\documentclass[11pt]{article}
\usepackage{amsmath,amssymb,amsfonts} 
\usepackage{graphics}                 
\usepackage{makeidx}
\usepackage{graphicx}
\usepackage{sidecap}
\usepackage[caption=false,font=normalsize,labelfont=sf,textfont=sf]{subfig}
\usepackage{algorithmic}
\usepackage[algoruled,vlined]{algorithm2e}
\usepackage{fullpage}
\usepackage{float}

\numberwithin{equation}{section}
\numberwithin{figure}{section}
\numberwithin{table}{section}

\newtheorem{theorem}{Theorem}[section]
\newtheorem{proposition}[theorem]{Proposition}
\newtheorem{corollary}[theorem]{Corollary}
\newtheorem{lemma}[theorem]{Lemma}

\newtheorem{definition}[theorem]{Definition}

\let \th=\theta

\let \rr=\rho
\let \k=\kappa
\let \n=\nu

\let \w=\omega

\let \la=\lambda

\let \o=\sigma
\def \a{\alpha}

\def \az{analytic}

\def \as{approximat}

\def \fr{filter}

\def \fb{filter bank}

\def \dn{denoising}

\def \sy{symmetr}
\def \df{differen}

\def \srr{\stackrel{\Delta}{=}}

\def \pa{\paragraph}

\def \v{vector}

\def \ww{wavelet}

\def \btt{\begin{theorem}}
\def \bll{\begin{lemma}}
\def \bdd{\begin{definition} }
\def \bpp{\begin{proposition} }
\def \bcc{\begin{corollary} }

\def \ett{\end{theorem}}
\def \ell{\end{lemma}}
\def \edd{\end{definition}}
\def \epp{\end{proposition}}
\def \ecc{\end{corollary} }

\def \we{waveform}

\def \pr{processing}

\def \on{orthonormal}
\def \oo{orthogonal}

\def \ry{represent}

\def \al{algorithm}

\def \cx{coefficient}
\def \fd{function}

\def \ff{frequency}

\def \s{spline}

\def \sp{spectr}
\def \de{direction}

\def \p{periodic}

\def \wq{wavelet packet}

\def \mv{modulation  matri}

\def \pl{polynomial}

\def \eh{Eq. \rf}

\def \d{decomposition}
\def \r{reconstruction}

\def \dt{discrete-time}

\def \t{transform}
\def \F{Fourier}

\newcommand{\rf}[1]{(\ref{#1})}

\newtheorem{thm}{Theorem}
\newtheorem{rmk}{Remark}[section]

\newtheorem{exm}{Example}

\def \srr{\stackrel{\mathrm{def}}{=}}

\def \bq{\begin{quote}}
\def \eq{\end{quote}}

\def \bt{\begin{thm}}
\def \et{\end{thm}}

\def \br{\begin{rmk}}
\def \er{\end{rmk}}

\def \bex{\begin{exm}}
\def \eex{\end{exm}}

\begin{document}
\title{\emph{Cross-boosting of WNNM Image Denoising method by Directional Wavelet Packets } }

\author{Amir Averbuch$^1$~~Pekka Neittaanm\"aki$^2$~~Valery  Zheludev$^1$~~Moshe Salhov$^1$ ~~Jonathan Hauser$^3$  \\
$^1$School of Computer Science, $^3$School of Electrical Engineering\\
Tel Aviv University, Israel\\
$^2$Faculty of Mathematical Information Technology\\
 University of Jyv\"askyl\"a, Finland}

 \date{ }
\maketitle
\begin{abstract}

The  paper presents an image denoising  scheme  by combining a method that is based on  directional quasi-analytic  wavelet packets (qWPs) with the state-of-the-art Weighted Nuclear Norm Minimization (WNNM) \dn\ algorithm.
The  qWP-based  denoising method (qWPdn) consists of  multiscale qWP \t\   of the degraded image, application of adaptive localized soft thresholding to the transform coefficients using the \emph{Bivariate Shrinkage} methodology,  and  restoration of  the image from the thresholded coefficients from several decomposition  levels. The combined method consists of several iterations of qWPdn and WNNM \al s in a way that at each iteration the output from one \al\  boosts the input to the other. The proposed methodology couples the qWPdn capabilities to capture edges and fine texture patterns even in the severely corrupted images with utilizing the  non-local self-similarity  in real images that is inherent in the WNNM \al .
 Multiple experiments, which    compared    the proposed methodology    with  six advanced \dn\ \al s, including WNNM,   confirmed that  the combined  cross-boosting \al\ outperformed most of them in terms of both quantitative measure and visual perception
quality.

\end{abstract}

\section{Introduction}\label{sec:s1}

High quality denoising is one of the main challenges in image \pr. It tries to achieve suppression of noise while capturing and preserving edges and fine structures in the image. A huge number of publications related to a variety of \dn\ methods (see, for example the reviews \cite{G0yRevDeno,deep_review,dn_cnn_rev}) exist.

Currently, two main groups of   image \dn\ methods  exist:\begin{enumerate}
                                                     \item ``Classical" schemes, which operate on  single images;
                                                     \item Methodologies based on Deep Learning.
                                                   \end{enumerate}
We briefly discuss the relation between these groups of methods in Section  \ref{sec:ss34}.

     Most up to date    ``classical" schemes,  where the proposed    \al\ belongs to,       are based on one of two approaches.

\begin{description}
  \item[Utilization of non-local self-similarity (NSS) in images:] Starting from the introduction of the Non-local mean (NLM) filter in \cite{nlm}, which is based on the similarity between pixels in \df t parts of the image, the exploitation of various forms of the NSS in images has resulted in a remarkable progress in image \dn . It is reflected in multiple publications (\cite{bm3d, sapca,wnnm,SABM1_3D,ncsr,Liu_osher,nlSS}, to name a very few). The  NSS  is explored in some \dn\ schemes based on Deep Learning (\cite{tnrd, dncnn}, for example).

A kind of benchmark in image \dn\ remains the BM3D \al\ ( \cite{bm3d}), which was presented as far as 2007. The \al\ exploits the self-similarity of patches  and  sparsity of the image in a \t\ domain. It collects similar patches in the image into a 3D array, which is subjected to a decorrelating 3D \t\ followed by either hard thresholding or Wiener \fr ing. After the inverse \t s, the processed patches  are returned to their original locations with corresponding weights. This method  is highly efficient  in restoration of moderately noised images.  However, the BM3D  tends to over-smooth and smear the image fine structure and edges when noise is strong.  

Some improvement of the original BM3D \al\ was achieved by using shape adaptive neighborhoods and the inclusion of  the Principal Component Analysis (PCA)  into the 3D \t\ (BM3D-SAPCA \al , \cite{sapca}). Even better results compared to BM3D and BM3D-SAPCA are demonstrated by the so-called Weighted Nuclear Norm Minimization (WNNM) method (\cite{wnnm}), which is based on the assumption that, by stacking the nonlocal
similar patch vectors into a matrix, this matrix should
be a low rank matrix and, as such, must  have sparse singular values. The  low rank matrix \as ion in \cite{wnnm} is achieved by an adaptive weighted thresholding  SVD values of such matrices.
Many \dn\ \al s presented in recent years, which are based on the  NSS concept, report results close to the results produced by BM3D-SAPCA and WNNM. At the same time, they share, to some extent, the shortcomings of the BM3D \al , especially blurring  the  fine structure of  images  restored from the  severely degraded inputs.
  \item[Transform domain \fr ing using \de al \fr s:] A way to capture lines, edges and texture pattern while restoring degraded images is to use \de ional \fr s and, respectively, dictionaries of \we s oriented in multiple \de s and having an oscillatory structure. A  number of dictionaries are reported in the literature and applied to image processing. We mention contourlets \cite{Contour}, curvelets \cite{curve,curve1},
 pseudo-polar Fourier transforms \cite{averbuch2008frameworkI,averbuch2008frameworkII} and related to them shearlets \cite{kuty,shear}. However, while these transforms successfully capture edges in images, these  dictionaries did not demonstrate a satisfactory  texture restoration due to  the shortage of oscillating waveforms in the dictionaries.

A number of publications \cite{king1,barakin,jalob1,bay_sele,bhan_zhao,bhan_com_sup, bhan_zhao_zhu, shenGabfr1,shenGabfr2}, to name a few,   derive  directional  dictionaries by the tensor multiplication of complex wavelets, wavelet  frames and wavelet packets (WPs).  The tight tensor-product complex  wavelet  frames (TP\_$\mathbb{C}$TF$_{n}$)\footnote{The index $n$  refers to the number of filters
in the underlying one-dimensional complex tight framelet filter bank.} with  different  numbers of  directions, are designed in \cite{bhan_zhao,bhan_zhao_zhu,bhan_com_sup} and some of them, in particular cptTP\_$\mathbb{C}$TF$_{6}$, TP\_$\mathbb{C}$TF$_{6}$ and TP\_$\mathbb{C}$TF$^{\downarrow}_{6}$, demonstrate impressive performance for image denoising and inpainting. The waveforms in these frames are oriented in 14 directions and, due to the 2-layer structure of their spectra, they possess certain, although  limited, oscillatory properties.

In
\cite{che_zhuang} (algorithm   \emph{Digital Affine Shear Filter Transform with 2-Layer Structure (DAS-2)})  the two-layer structure, which is inherent in the TP\_$\mathbb{C}$TF$_{6}$  frames,  is incorporated into shearlet-based directional filter banks introduced in \cite{zhuang}. This improves the  performance of DAS-2 in comparison to TP\_$\mathbb{C}$TF$_{6}$ on texture-rich images, which is not the case for smoother images.

Recently, we  designed a
family of complex WPs  (\cite{azn_pswq}, brief outlook of the design is in \cite{azn_Impwq}), which are referred to as quasi-\az\ WPs (qWPs). As a base for the design,  the family of  WPs originated from \p\ \s s of different orders, which are described in \cite{ANZ_book3} (Chapter 4), is used. The two-dimensional (2D) qWPs are derived by a standard tensor products of 1D qWPs. The real parts of the 2D qWPs possess a combination of properties valuable for  image \pr : They are oriented in multiple \de s, (see Table \ref{dir_count});  The \we s are close to  directional cosine waves with multiple  frequencies modulated  by  localized low-\ff\ 2D  signals; Their DFT \sp a form a refined split of the \ff\ domain;
Both one- and two-dimensional qWP \t s are implemented in a very  fast ways by using the Fast \F\ \t\ (FFT). The \de al qWPs are successfully applied to image inpainting (\cite{azn_Impwq}).
\end{description}

Due to the above properties, a  qWP-based  \dn\ \al\  (qWPdn), which utilizes  an adapted version of the Bivariate Shrinkage \al\ (BSA \cite{bishr,sen_seles}), proved to be efficient for image \dn.
Experiments with the qWPdn demonstrate its  ability to restore edges and texture details even from severely degraded images.   In most experiments, the qWPdn to be described in Section \ref{sec:ss31} provides better resolution of edges and fine structures compared to the cptTP-$\mathbb{C}$TF$_6$, DAS-2 and NSS-based \al s,  which is reflected in getting higher Structural Similarity Index (SSIM)\footnote{\cite{ssim}, \texttt{ssim.m} Matlab 2020b \fd .} values. On the other hand, the NSS-based  \al s, especially  WNNM,  proved to be superior in the noise suppression, especially  in smooth regions of images, thus producing the highest PSNR values in almost all the experiments. However, some  over-smoothing effect on  the edges and fine texture persisted under the BM3D, BM3D-SAPCA, NCSR (\cite{ncsr})   and WNNM \al s.
Especially, this is the case for severely degraded images.

Therefore, we propose  to combine the  qWPdn and  WNNM \al s in order to retain strong features of both \al s and to get rid of their drawbacks. The  combined   qWPdn-- WNNM \al s presented in the paper consist of the iterated execution of  the  qWPdn and  WNNM \al s  in a  way that at each iteration, the output from one \al\ updates (boosts) the input to the other. Typically, 2--3  (rarely more than 4) iterations are needed to get an excellent result.

In multiple  experiments, part of which is reported in Section \ref{sec:ss33},  the qWPdn--WNNM  \al s  performance is compared with the    performance of the NSS-based  BM3D, BM3D-SAPCA, NCSR and WNNM \al s and the \al s cptTP-$\mathbb{C}$TF$_6$ and  DAS-2 using \de al \fr s.  The  hybrid  qWPdn--WNNM \al s demonstrated  noise suppression efficiency that is quite competitive with   all the above methods. It produces PSNR values higher than BM3D produces and either very close to or higher than the values produced by the BM3D-SAPCA, NCSR and WNNM \al s. On the other hand, its performance related to the edge   and fine structures resolution is much better than the performance of all the participated   \al s, thus, producing significantly higher SSIM values.

\begin{figure}
\centering
\includegraphics[width=3.2in]{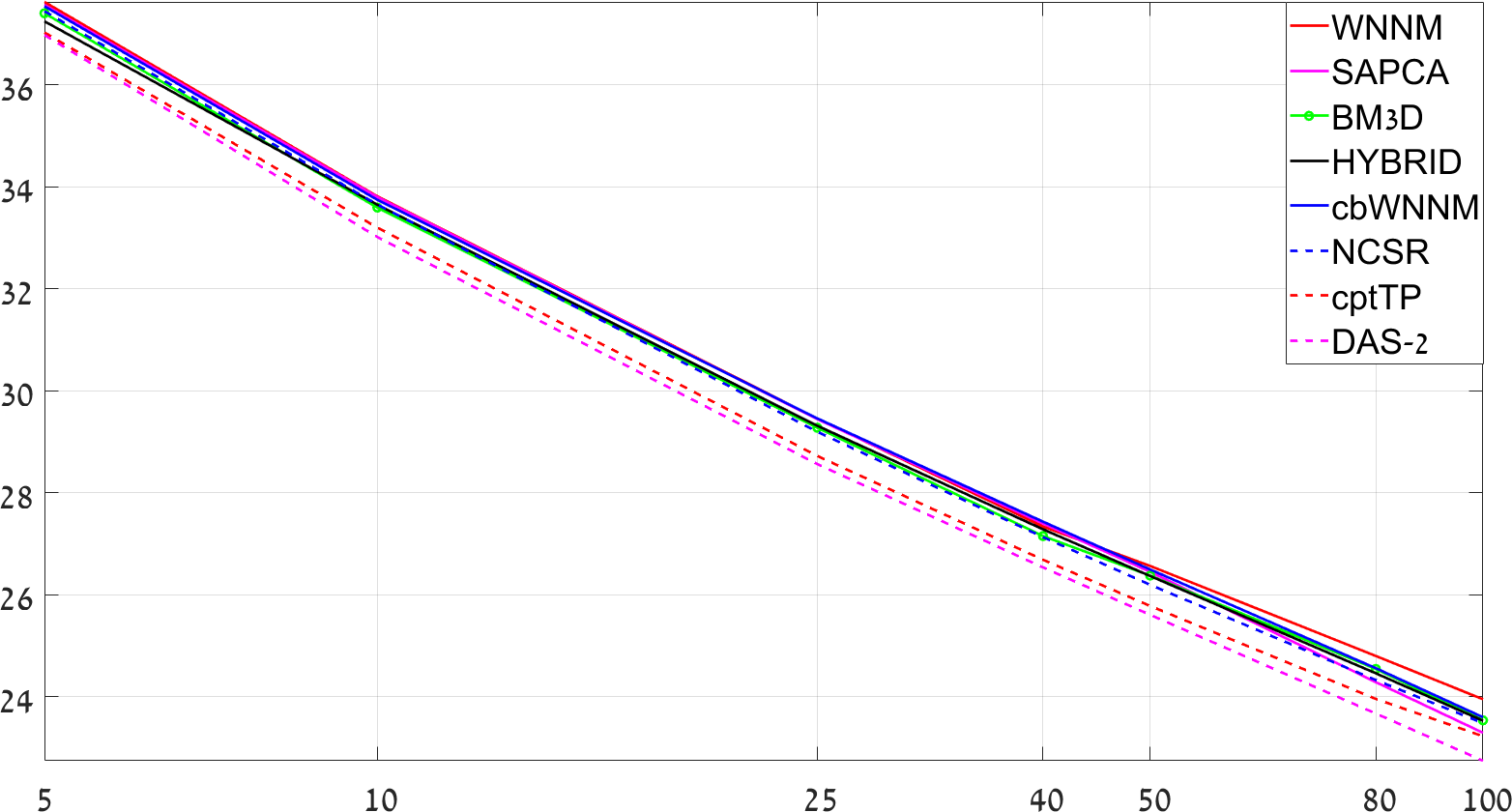}%
\includegraphics[width=3.2in]{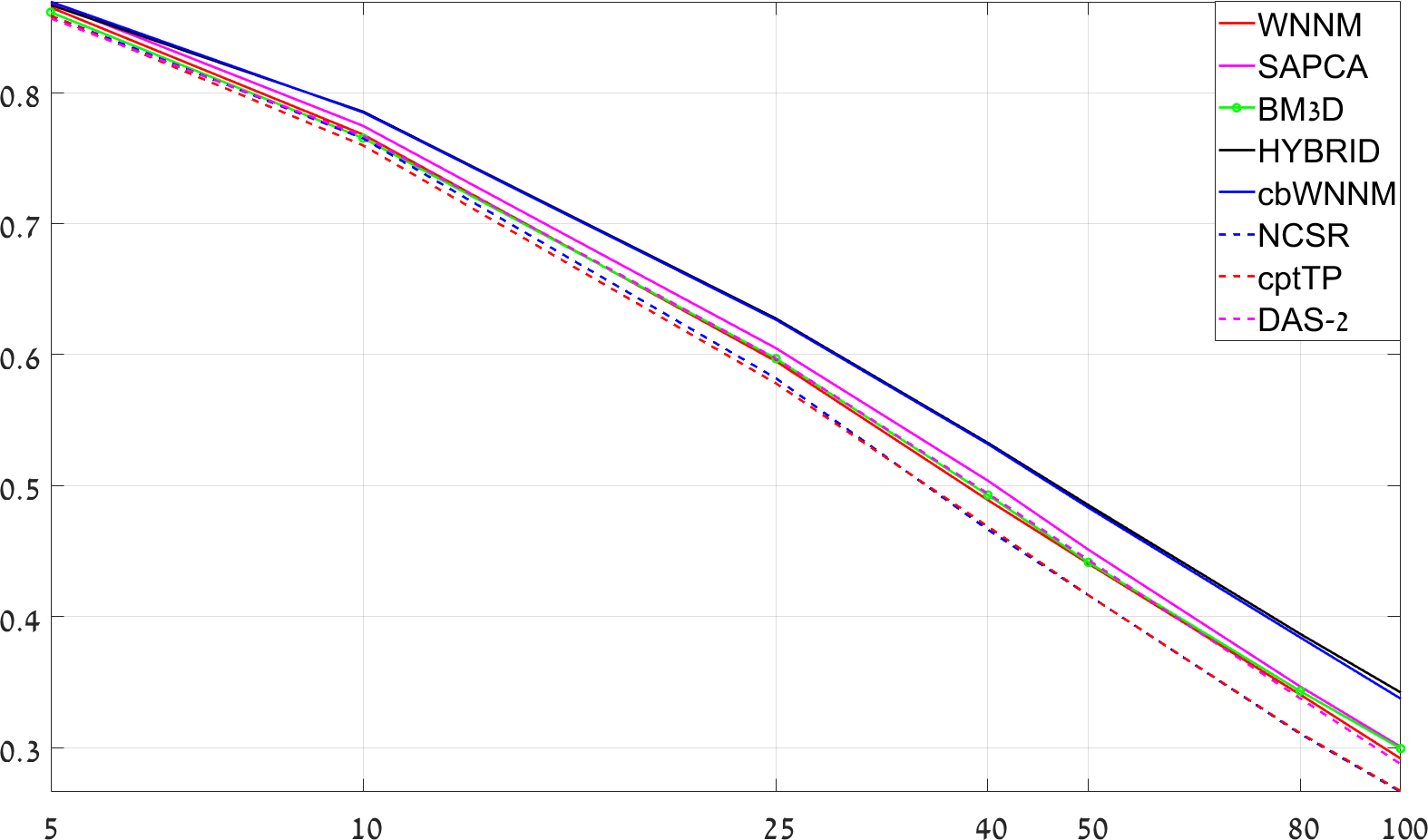}%
\caption{Diagrams of  PSNR (left frame) and SSIM (right frame) values averaged over ten images for restoration of the images  by eight \df t methods. Labels on \textbf{x}-axes indicate STD=$\sigma$  values of the additive noise (logarithmic scale)}
\label{diaPS}
\end{figure}

This observation is illustrated by diagrams in Fig. \ref{diaPS}. The left
 frame in Fig. \ref{diaPS} shows PSNR values for the restoration of images degraded by Gaussian noise with STD $\sigma=$ 5, 10, 25, 40 50, 80 and 100 dB, by  the above   six  methods and two combined
 qWPdn--WNNM  \al s designated by \textbf{cbWNNM} and {\textbf{hybrid}}. The  PSNR values are averaged over ten images participated in the experiments (see Fig. \ref{clima}). The right  frame in  Fig.  \ref{diaPS}  does the same for the SSIM values. We can observe that the averaged PSNR values for our methods  \textbf{cbWNNM} and {\textbf{hybrid}}  practically coincide with each other and are very close  to the values produced by BM3D-SAPCA and WNNM.
A different situation we see in the right  frame, which displays averaged  SSIM values. Again,  the averaged values for our methods  \textbf{cbWNNM} and {\textbf{hybrid}}  practically coincide with each other but they strongly override the values produced by all other  methods.

\paragraph{Contribution of the paper:}
\begin{itemize}
  \item Development  of fast image \dn\ \al\ qWPdn based on recently designed directional quasi-\az\ \wq s.
  \item Design of the  iterative cross-boosting qWPdn--WNNM \al s, which are highly efficient in noise suppression and capturing edges and fine structures even in severely degraded images.
  \item Experimental comparison of the qWPdn--WNNM \al s performance with the performance  of multiple state-of-the-art \al s, which demonstrated a decisive advantage of these\al s in a SSIM sense and visual perception.
\end{itemize}

The paper is organised as follows:
Section \ref{sec:s2}  briefly outlines properties  of the  qWP \t s in one and two dimensions.  Section \ref{sec:ss31} describes  the qWPdn \al.  Section \ref{sec:ss32} presents  the combined  qWPdn--WNNM \al s. In the multiple experiments in Section \ref{sec:ss33}, the performance of these \al s  is  compared with the performance of the cptTP-$\mathbb{C}$TF$_6$, DAS-2,  BM3D, BM3D-SAPCA, NCSR and WNNM \al s. Section \ref{sec:ss34} briefly discusses the  relation of the proposed methodology  to the recently published  Deep Learning denoising  methods.  Section \ref{sec:s4} provides an overview of the results.

\paragraph{Notation and abbreviations:}
$N=2^{j}$,  $\w\srr e^{2\pi\,i/N}$ and $\Pi[N]$ is a space of real-valued  $N$-\p\ signals.
$\Pi[N,N]$ is the space of two-dimensional   $N$-\p\  in both vertical and horizontal directions arrays.
DFT(FFT) means Discrete(Fast) \F\ \t .

The abbreviations WP, dWP and qWP mean \wq,  \on\ \s-based \wq\ $\psi^{p}_{[m],l}$ and quasi-\az\ \wq s $\Psi^{p}_{\pm[m],l}$, respectively, in a 1D case, and  \on\ WPs $\psi^{p}_{[m],j,l}$ and quasi-\az\ \wq s $\Psi^{p}_{+\pm[m],l,j}$, respectively, in a 2D case.

   qWPdn designates the qWP-based image \dn\ \al. qWPdn--WNNM means a cross-boosting image \dn\ \al\ combining the qWPdn with the BM3D.

PSNR means Peak Signal-to-Noise ratio in decibels (dB).
SSIM means Structural Similarity Index (\cite{ssim}) computed by the Matlab 2020b \fd\ \texttt{ssim.m}.
BSA stands for Bivariate Shrinkage \al\  (\cite{sen_seles,bishr}).
NSS means non-local self-similarity.

BM3D stands for \emph{Block-matching and 3D filtering}  (\cite{bm3d}), SAPCA means Shape-Adaptive Principal Component Analysis (\cite{sapca}), NCSR means Nonlocally Centralized Sparse Representation  (\cite{ncsr}),  WNNM means Weighted Nuclear Norm Minimization (\cite{wnnm}), cptTP-$\mathbb{C}$TF stands for
 \emph{Compactly Supported Tensor Product Complex
Tight Framelets with Directionality}  (\cite{Zhu_han})  and DAS-2 stands for \emph{Digital Affine Shear Filter Transform with 2-Layer Structure} (\cite{che_zhuang}).

\section{Preliminaries: Quasi-analytic \de al \wq s}\label{sec:s2}
Recently we designed a family of  quasi-analytic \wq s (qWPs), which possess a  collection of properties indispensable for image \pr. A brief outline of the qWPs design and the implementation of corresponding \t s is provided  the paper \cite{azn_Impwq}, which describes successful application of qWPs to  image inpainting. A detailed description of the design and  implementation is given in \cite{azn_pswq}.
In this section we list  properties of qWPs and  present some illustrations.

\subsection{Properties of qWPs}\label{sec:ss21}
\pa{One-dimentional qWPs} The qWPs  are derived  from the  \p\  WPs originating from  \on\  discretized  \pl\ \s s of \df t orders (dWPs), which are described in  Chapter 4 in \cite{ANZ_book3} (a brief outline is given in \cite{azn_pswq}).
The dWPs are denoted by $\psi^{p}_{[m],l}$, where $p$ means the generating \s's order, $m$ is the \d\ level and $l=0,...2^m-1,$ is the index of an $m$-level \wq s. The $2^{m}$-sample shifts $\left\{\psi^{p}_{[m],l}(\cdot -2^{m}\,k)\right\},\;l=0,...,2^m-1,\;k=0,...,N/2^m-1,$ of the $m$-level dWPs form an \on\ basis of the space $\Pi[N]$ of $N$-\p\ \dt\ signals. Surely, other \on\ bases are possible, for example, \ww\ and Best bases (\cite{coiw1}).

The \we s $\psi^{p}_{[m],l}[k]$ are \sy ic, well localized in the spatial domain and have oscillatory structure. Their DFT \sp a form a refined split of the \ff\ domain. The shapes of magnitude \sp a  tend to rectangular as the \s's order $p$ grows.
A common way to extend  1D WP \t s  to multiple dimensions is by the tensor-product extension. The 2D  dWPs from the level $m$ are: $\psi_{[m],j ,l}^{p}[k,n]\srr\psi_{[m],j}^{p}[k]\,\psi_{[m],l}^{p}[n]$. Their $2^{m}$-sample shifts along vertical and horizontal directions form \on\ bases of the space $\Pi[N,N]$ of 2D signals $N-$\p\ in both directions. The drawback for image \pr\ is the lack of \de ality. The \de ality can be achieved by switching to complex \wq s.

For this, we start with application of the Hilbert \t\ (HT) to the dWPs  $\psi^{p}_{[m],l}$, thus getting the signals  $\tau^{p}_{[m],l}=H(\psi^{p}_{[m],l}), \; m=1,...,M,\;l=0,...,2^m-1$. A slight correction of   those signals \sp a:  \begin{equation}\label{cwq}
  \hat{\phi}^{p}_{[m],l}[n]\srr \hat{\psi}^{p}_{[m],l}[0]+\hat{\psi}^{p}_{[m],l}[N/2]+   \hat{\tau}^{p}_{[m],l}[n]
  \end{equation}
  provides us with a set of signals from the space $\Pi[N]$, whose properties are similar to the   properties of the dWPs  $\psi^{p}_{[m],l}$. In particular, their shifts form \on\ bases in $\Pi[N]$, their magnitude \sp a coincide with the magnitude \sp a of the dWPs  $\psi^{p}_{[m],l}$. However, unlike  the \sy ic dWPs  $\psi^{p}_{[m],l}$, the signals  $\phi^{p}_{[m],l}$ are anti\sy ic for all $l$ except for $l_{0}=0$ and $l_{m}=2^{m}-1$.
We refer to the signals  $\phi^{p}_{[m],l}$ as the complementary \on\  WPs (cWPs).

The sets of complex-valued WPs, which we refer to as the quasi-\az\  wavelet packets (qWP),  are defined as
 $ \Psi^{p}_{\pm[m],l}\srr\psi^{p}_{[m],l}   \pm i\phi^{p}_{[m],l}, \quad m=1,...,M,\;l=0,...,2^{m}-1$,
where $\phi^{p}_{[m],l}$ are the cWPs defined in \eh{cwq}. The qWPs $\Psi^{p}_{\pm[m],l}$ differ from the \az\ WPs  by adding  two values $\pm i\,\hat{\psi}^{p}_{[m],l}[0]$ and $\pm i\,\hat{\psi}^{p}_{[m],l}[N/2]$ into their DFT \sp a, respectively.
The DFT \sp a  of the qWPs $ \Psi^{p}_{+[m],l}$ are located within positive half-band of the \ff\ domain and vice versa for the qWPs $ \Psi^{p}_{-[m],l}$.

\begin{figure}
\begin{center}
\includegraphics[width=6in]{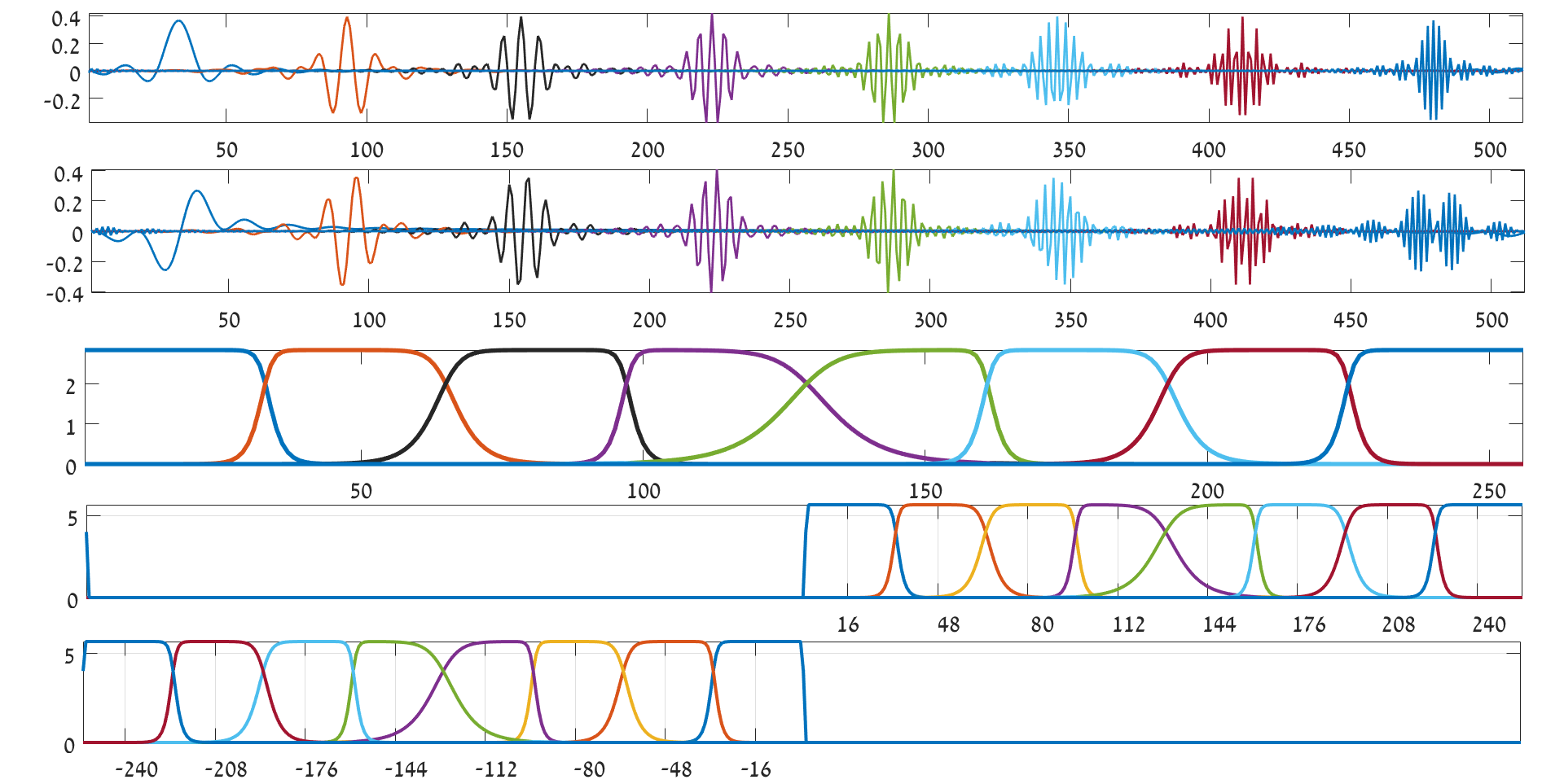}
\end{center}
\caption{Top to bottom: signals ${\psi}^{9}_{[3],l}$; signals ${\phi}^{9}_{[3],l},\;l=0,...,7$:  their magnitude DFT spectrum  (right half-band); magnitude DFT spectra of complex qWPs ${\Psi}^{9}_{+[3],l}$; same for ${\Psi}^{9}_{-[3],l},\;l=0,...,7$}
\label{psi_phiT}
\end{figure}

 Figure \ref{psi_phiT} displays  the signals ${\psi}^{9}_{[3],l}$ and ${\phi}^{9}_{[3],l},\;l=0,...,7$, from the third decomposition  level and their magnitude spectra (right half-band),  that coincide with each other.  Addition of $\hat{\psi}^{9}_{[3],l}[0]$ and $\hat{\psi}^{9}_{[3],l}[N/2]$ to the  spectra of ${\phi}^{9}_{[3],l},\;l=0,7,$ results in an antisymmetry distortion.  These WPs provide a collection of diverse \sy ic  and anti\sy ic  well localized \we s, which range from smooth \ww s for $l=0,1$ to fast oscillating transients for $l=5,6,7$. Thus,  this collection is well suited to catching smooth as well as  oscillating local  patterns in signals. In the 2D case, these valuable properties of the \s-based \wq s are completed by the directionality of the tensor-product \we s.

\pa{Two-dimensional qWPs}
 Similarly to the 2D dWPs  $\psi_{[m],j ,l}^{p}[k,n]$, the 2D cWPs  $\phi_{[m],j ,l}^{p}[k,n]$ are defined as the tensor products of 1D WPs such that
\(
 \phi_{[m],j ,l}^{p}[k,n]\srr\phi_{[m],j}^{p}[k]\,\phi_{[m], l}^{p}[n].
\)
The $2^{m}$-sample shifts of the WPs $\left\{\phi_{[m],j ,l}^{p}\right\},\;j , l=0,...,2^{m}-1,$ in both directions form an \on\ basis for the space $\Pi[N,N]$ of arrays that are $N$-\p\ in both directions.

\pa{2D qWPs and their \sp a }
The 2D dWPs  $\left\{\psi_{[m],j ,l}^{p}\right\}$ as well as the cWPs  $\left\{\phi_{[m],j ,l}^{p}\right\}$  lack  the directionality property  which is needed in many applications that process 2D data.  However, real-valued 2D \wq s oriented in multiple directions  can be
derived from  tensor  products of complex quasi-\az\ qWPs $\Psi_{\pm[m],\rr}^{p}$.
The complex 2D qWPs are defined  as follows:
\begin{eqnarray} \label{qwp_2d}
\Psi_{++[m],j , l}^{p}[k,n] \srr \Psi_{+[m],j}^{p}[k]\,\Psi_{+[m], l}^{p}[n], \quad
  \Psi_{+-[m],j ,l}^{p}[k,n] \srr\Psi_{+[m],j}^{p}[k]\,\Psi_{-[m], l}^{p}[n],
\end{eqnarray}
where $  m=1,...,M,\;j ,l=0,...,2^{m}-1,$ and $k ,n=0,...,N-1$.
The real  parts of these 2D qWPs are
\begin{equation}
\label{vt_pm}
\begin{array}{lll}
 \th_{+[m],j ,l}^{p}[k,n] &\srr& \mathfrak{Re}(\Psi_{++[m],j ,l}^{p}[k,n]) =  \psi_{[m],j ,l}^{p}[k,n]-\phi_{[m],j ,l}^{p}[k,n], \\
\th_{-[m],j ,l}^{p}[k,n] &\srr&  \mathfrak{Re}(\Psi_{+-[m],j ,l}^{p}[k,n]) =  \psi_{[m],j ,l}^{p}[k,n]+\phi_{[m],j ,l}^{p}[k,n].\\
\end{array}
\end{equation}

\begin{SCfigure}
\centering
\caption{Block-scheme of the qWP design (left) and quadrants of  frequency  domain (right)}
\includegraphics[width=2.1in]{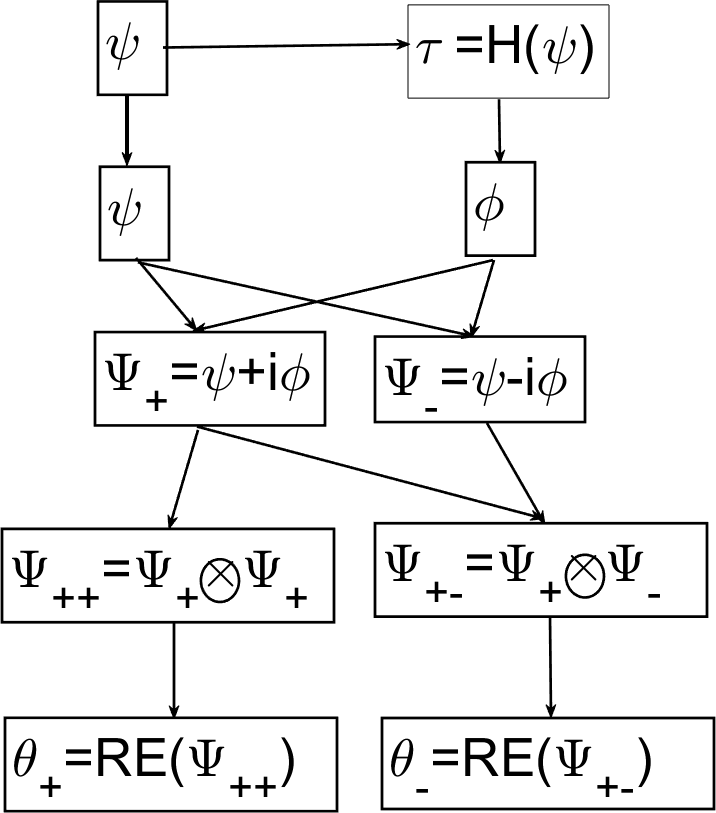}\quad \vline\quad  \includegraphics[width=1.4in]{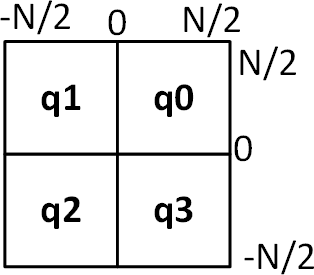}
  \label{dia_fipsi}
\end{SCfigure}

\begin{SCfigure}
\centering
\caption{Magnitude \sp a of 2D qWPs $\Psi_{++[2],j ,l}^{9}$ (left) and $\Psi_{+-[2],j ,l}^{9}$  (right) from the second \d\ level}
\includegraphics[width=2in]{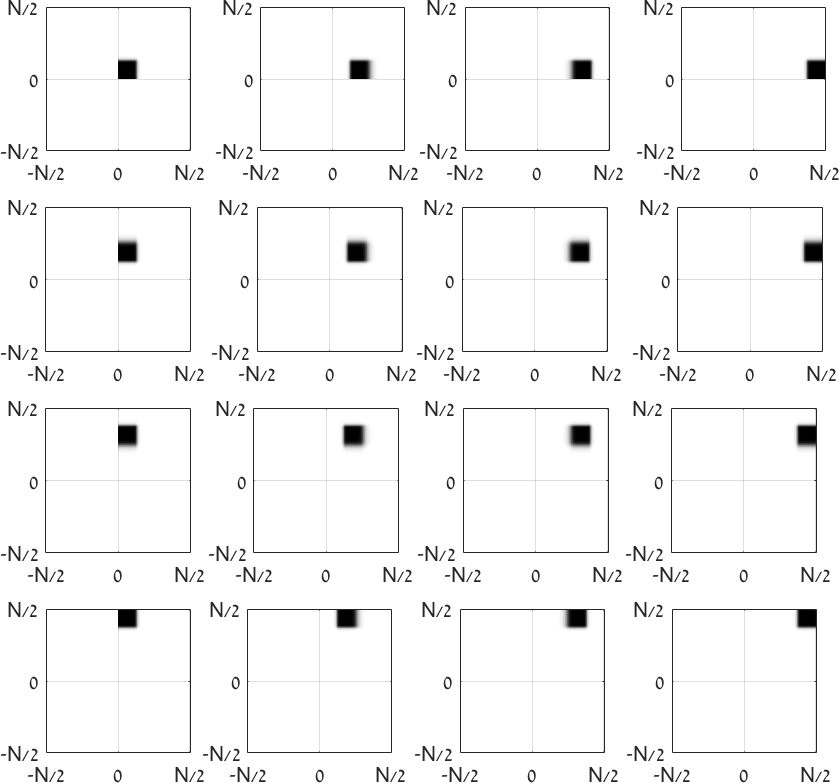}
\quad\vline\quad
\includegraphics[width=2in]{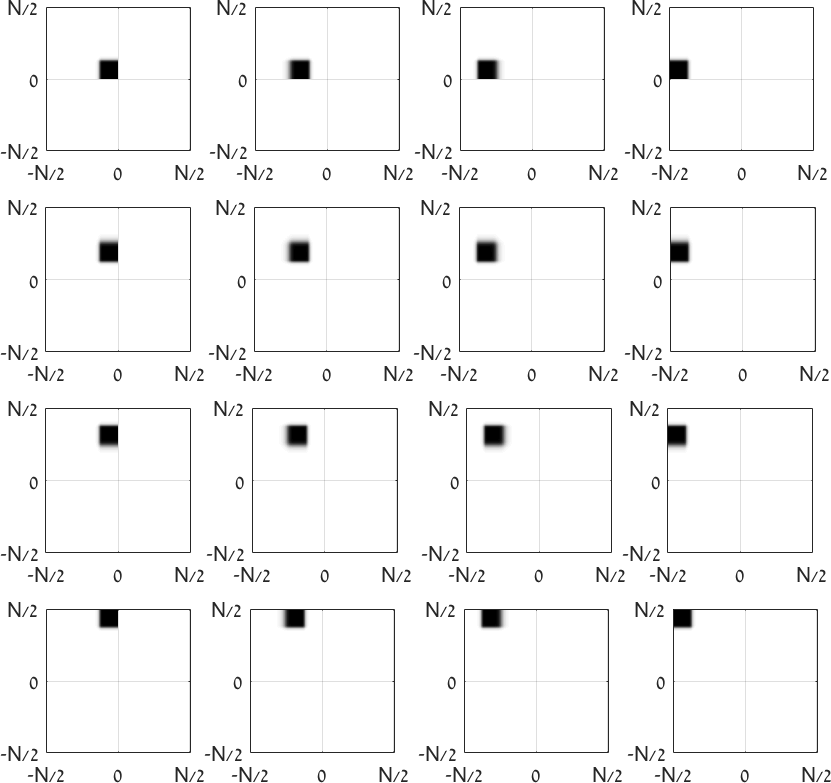}%
\label{fpp_2}
\end{SCfigure}

\begin{SCfigure}
\centering
\caption{Left: magnitude \sp um of 2D qWP $  \Psi_{++[3],2 ,5}^{p}[k,n]$. Right    WP $ \th_{++[3],2 ,5}^{p}=\mathfrak{Re}( \Psi_{++[3],2 ,5}^{p})$}
\includegraphics[width=2.6in]{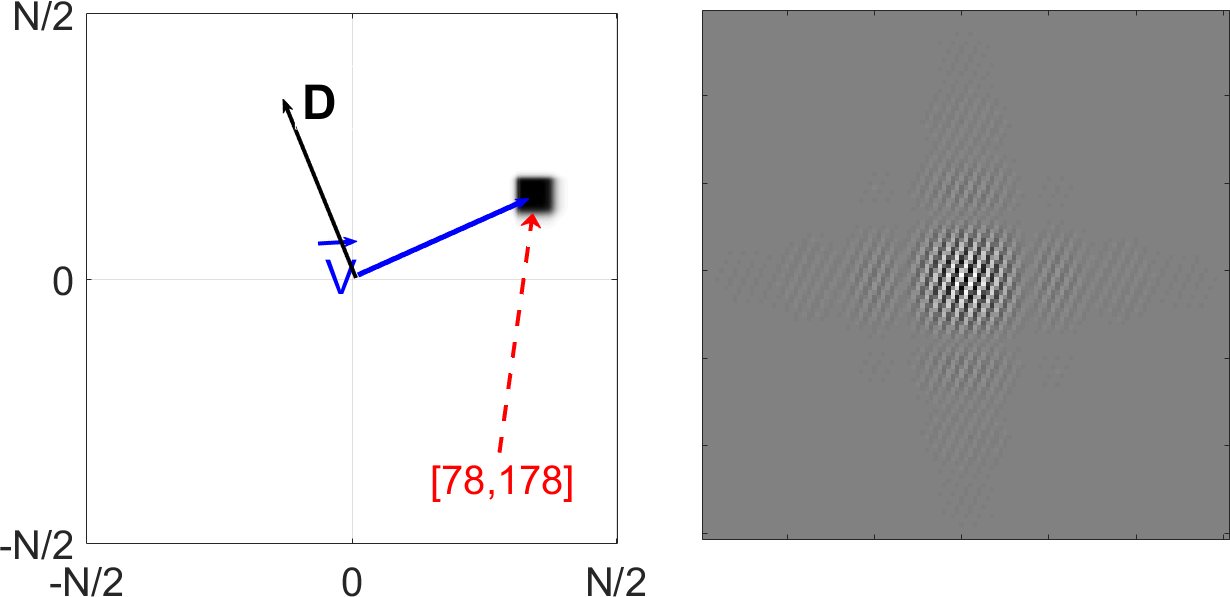}
\label{78_178}
\end{SCfigure}

The block-scheme in Fig. \ref{dia_fipsi} illustrates the design of  qWPs.
The DFT \sp a of the 2D qWPs $\Psi_{++[m],j ,l}^{p},\;j ,l=0,...,2^{m}-1,$ are  tensor products of the one-sided \sp a of the qWPs
$\hat{ \Psi}_{++[m],j ,l}^{p}[p,q] =\hat{ \Psi}_{+[m],j}^{p}[p]\,\hat{\Psi}_{+[m], l}^{p}[q]$
and, as such,  they fill the  quadrant $\mathbf{q}_{0} $ of the \ff\ domain, while the \sp a of $\Psi_{+-[m],j ,l}^{p},\;j ,l=0,...,2^{m}-1,$ fill the  quadrant $\mathbf{q}_{1}$ (see Fig. \ref{dia_fipsi}).
Figure \ref{fpp_2} displays magnitude \sp a of the ninth-order 2D qWPs $\Psi_{++[2],j ,l}^{9}$ and $\Psi_{+-[2],j ,l}^{9}$ from the second \d\ level.
Figure \ref{fpp_2} shows that the DFT \sp a of the  qWPs $\Psi_{+\pm[m],j ,l}^{9}$ effectively occupy relatively small squares in the \ff\ domain. For deeper \d\ levels,  sizes of the corresponding  squares decrease as geometric progression. Such configuration of the \sp a leads to the directionality of the real-valued 2D WPs $ \th_{\pm[m],j ,l}^{p}$. The directionality of the WPs $ \th_{\pm[m],j ,l}^{p}$ is discussed in \cite{azn_pswq}. It is established  that if the \sp um of a WP $\Psi_{+\pm[m],j ,l}^{p}$ occupies a square whose center lies in  the point $[\k_{0},\n_{0}]$, then the respective real-valued  WP $ \th_{\pm[m],j ,l}^{p}$ defined in \eh{vt_pm} is \ry ed by
\(  \th_{\pm[m],j ,l}^{p}[k,n]  \approx{\cos\frac{2\pi(\k_{0}k+\n_{0}n)}{N}}\,\underline{\th}[k,n] ,
\)
where $\underline{\th}[k,n]$ is a spatially localized low-\ff\ \we\ which does not have a directionality. But the 2D signal  $\cos\frac{2\pi(\k_{0}k+\n_{0}n)}{N}$ is oscillating in the direction \textbf{D}, which is \oo\ to  the vector $\vec{V}=\k_{0}\vec{i}+\n_{0}\vec{j}$. Therefore, WP $\th_{\pm[m],j ,l}^{p}$  can be regarded as the directional cosine wave modulated  by the localized low-\ff\ signal  $\underline{\th}$. The cosine frequencies in the vertical and horizontal directions are determined by the indices $j$ and $l$, respectively,  of the WP $ \th_{\pm[m],j ,l}^{p}$. The bigger is the index, the higher is \ff\ in the respective direction. The situation is illustrated in Fig. \ref{78_178}. The imaginary parts of the qWPs $\Psi_{+\pm[m],j ,l}^{p}$ have a similar structure.

\begin{figure}
\begin{center}
\includegraphics[width=3.in]{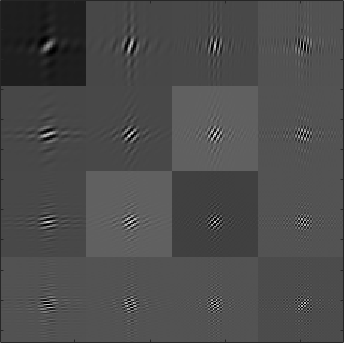}
\quad
\includegraphics[width=3.1in]{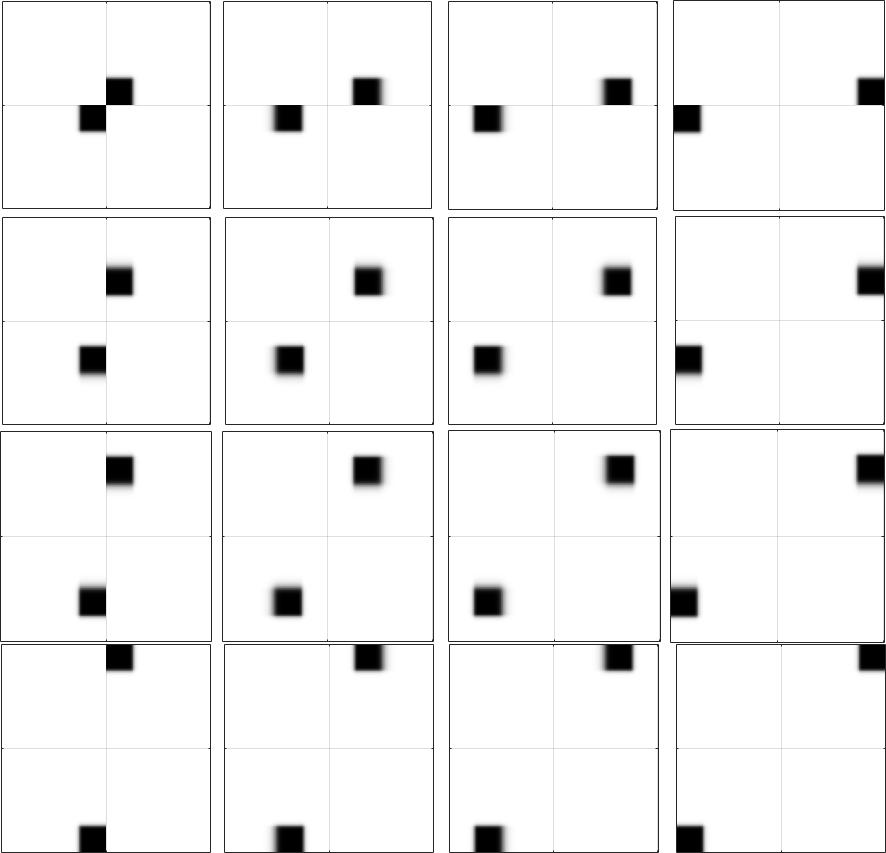}%
\end{center}
\centering
\caption{WPs $\th_{+[2],j ,l}^{9}$ from the second decomposition  level and their magnitude spectra}
\label{pp_2_2d}
\end{figure}

Figure \ref{pp_2_2d} displays WPs  $\th_{+[2],j ,l}^{9},\;j,l=0,1,2,3,$ from the second decomposition  level and their magnitude spectra.

\begin{figure}
\begin{center}
\includegraphics[width=2.9in]{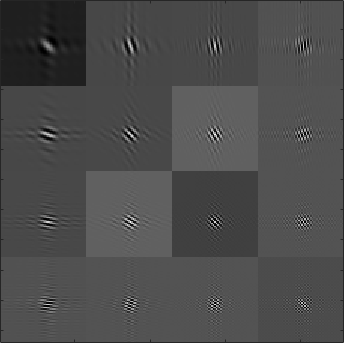}
\quad
\includegraphics[width=3.1in]{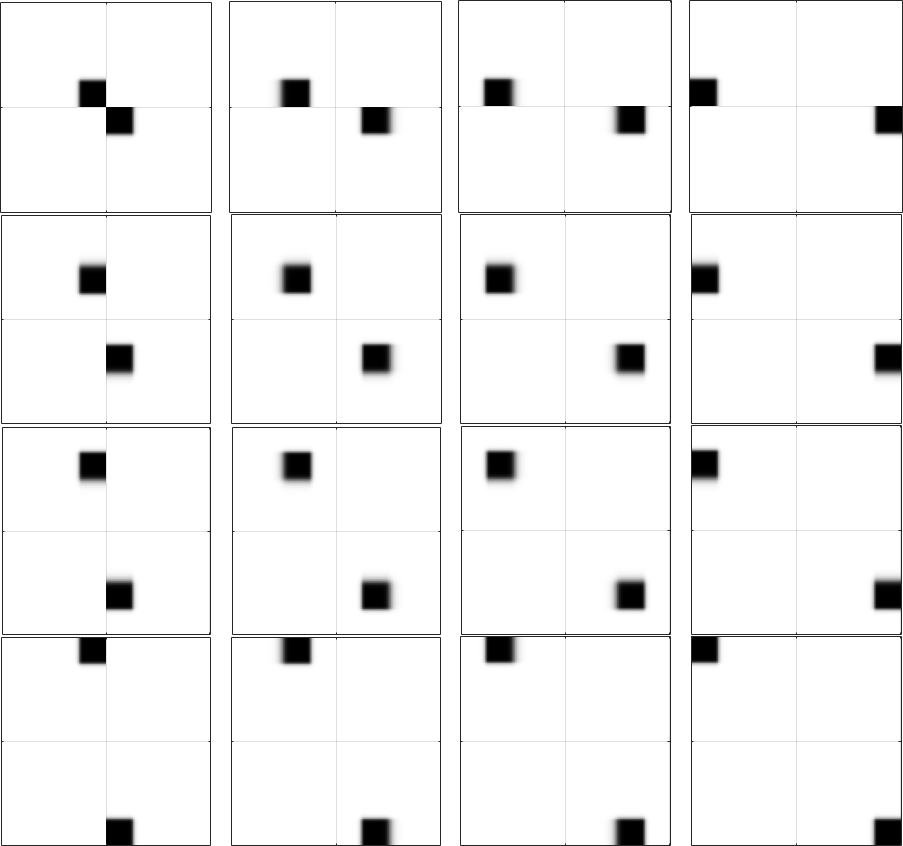}\centering
\end{center}
\caption{WPs $\th_{-[2],j ,l}^{9}$ from the second decomposition  level and their magnitude spectra}
\label{pm_2_2d}
\end{figure}

Figure \ref{pm_2_2d} displays WPs  $\th_{-[2],j ,l}^{9},\;j,l=0,1,2,3,$ from the second decomposition  level and their magnitude spectra.

\begin{figure}
\begin{center}
\includegraphics[width=3.in]{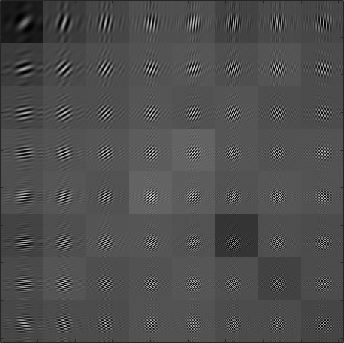}\quad\vline\quad
\includegraphics[width=3.in]{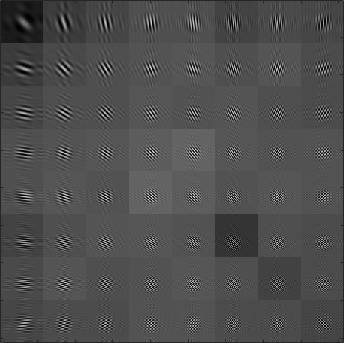}
\end{center}
\centering
\caption{WPs $\th_{+[3],j ,l}^{9}$  (left) and $\th_{-[3],j ,l}^{9}$  (right) from the third decomposition  level }
\label{pp_3_2d}
\end{figure}

Figure \ref{pp_3_2d} displays WPs $\th_{+[3],j ,l}^{9}$ and $\th_{-[3],j ,l}^{9},\;j,l=0,1,2,3,$ from the third decomposition  level.

\br\label{direc_rem}Note that all the  WPs $\th_{+[m],j ,l}^{p}$ whose \sp a are located along the \v\ $\vec{V}$ have \as ely the same orientation.  It is seen in Figs. \ref{pm_2_2d}, \ref{pp_2_2d} and \ref{pp_3_2d}. Consequently, the number of orientations of the  $m$-th level WPs is less than the number of WPs, which is $2\cdot4^{m}$. For example, all the ``diagonal" qWPs $\left\{\th_{\pm[m],j ,j}^{p}\right\},\;j=0,...,2^{m}-1,$ are oscillating with \df t frequencies in the directions of either $135^{o}$  (for $ \th_{+}$) or $45^{o}$  (for $ \th_{-}$).  Orientation  numbers are given in Table \ref{dir_count}.\er
\begin{table}
  \centering
  \caption{ Numbers of \df t orientations of  qWPs  $\left\{\th_{\pm[m],j ,l}^{p}\right\},\;j,l=0,...,2^{m}-1$, for \df t \d\ levels}\label{dir_count}
 \begin{tabular}{|l|l|l|l|l|l|l|l|}
     \hline
     level $m$& 1 & 2 & 3 & 4 & 5 & 6  & ... \\
     \hline
     \# of directions & 6 & 22 & 86 & 318 & 1290 & 5030 & ... \\
     \hline
   \end{tabular}
  \end{table}

\subsection{Outline of the implementation scheme for  2D qWP \t s}\label{sec:ss22}
 The \sp a of 2D qWPs $\left\{\Psi_{++[m],j ,l}^{p}\right\},\;j ,l=0,...,2^{m}-1$ fill the quadrant  $\mathbf{q}_{0}$  of the \ff\ domain (see Fig. \ref{dia_fipsi}), while  the \sp a of 2D qWPs $\left\{\Psi_{+-[m],j ,l}^{p}\right\}$ fill the quadrant  $\mathbf{q}_{1}$.
Consequently, the \sp a of the real-valued 2D WPs $\left\{\th_{+[m],j ,l}^{p}\right\},\;j ,l=0,...,2^{m}-1$, and  $\left\{\th_{-[m],j ,l}^{p}\right\}$ fill the pairs  of quadrant  $\mathbf{q}_{+}=\mathbf{q}_{0}\bigcup\mathbf{q}_{2}$ and $\mathbf{q}_{-}=\mathbf{q}_{1}\bigcup\mathbf{q}_{3}$, respectively.

By this reason, none linear combination of the WPs $\left\{\th_{+[m],j ,l}^{p}\right\}$  and their shifts can serve as a basis for  the signal  space $\Pi[N,N]$. The same is   true for WPs $\left\{\th_{-[m],j ,l}^{p}\right\}$. However,  combinations of  the WPs $\left\{\th_{+[m],j ,l}^{p}\right\}$ and $\left\{\th_{-[m],j ,l}^{p}\right\}$ provide frames of the  space $\Pi[N,N]$.

The \t s are implemented in the \ff\ domain using \mv ces of the \fb s, which are built from the corresponding \wq s. It is important to mention that the structure of the \fb s $\mathbf{Q}_{+}$ and $\mathbf{Q}_{-}$ for the first \d\ level is \df t for the \t s with the ``positive" $\Psi_{+[m],l}^{p}$ and ``negative" $\Psi_{-[m],l}^{p}$ qWPs, respectively. However,  the \t s from the first to the second and further \d\ levels are executed using the same \fb\ $\mathbf{H}_{m}$ for the ``positive" and ``negative" qWPs. This fact makes it possible a parallel implementation of the \t s.

The one-level 2D qWP \t s of a signal  $\mathbf{X}=\left\{X[k,n] \right\}\in\Pi[N,N]$ are implemented by a  tensor-product scheme. To be specific,
for the \t\ with $\Psi^{p}_{++[1]}$, the 1D \t\ of rows  from the  signal  $\mathbf{X}$ is executed using the \fb\ $\mathbf{Q}_{+}$, which is followed by the 1D \t\ of columns  of the produced coefficient  arrays  using the same \fb\ $\mathbf{Q}_{+}$. These operations  result in the \t\ coefficient  array $\mathbf{Z}_{+[1]}=\bigcup_{\j,l=0}^{1}\mathbf{Z}_{+[1]}^{j,l}$ comprising of four blocks of size $N/2\times N/2$.
  The \t\ with $\Psi^{p}_{+-[1]}$ is implemented by the subsequent application of the \fb s  $\mathbf{Q}_{+}$ and  $\mathbf{Q}_{-}$ to rows from the  signal  $\mathbf{X}$ and columns  of the produced coefficient  arrays, respectively. This  results in the coefficient   array $\mathbf{Z}_{-[1]}=\bigcup_{\j,l=0}^{1}\mathbf{Z}_{-[1]}^{j,l}$,

The further \t s starting from the arrays $\mathbf{Z}_{+[1]}$ and $\mathbf{Z}_{-[1]}$   produce two sets of the \cx s $\left\{\mathbf{Z}_{+[m]}=\bigcup_{\j,l=0}^{2^{m}-1}\mathbf{Z}_{+[m]}^{j,l}\right\}$ and $\left\{\mathbf{Z}_{-[m]}=\bigcup_{\j,l=0}^{2^{m}-1}\mathbf{Z}_{-[m]}^{j,l}\right\},\;m=2,...,M$.
The \t s are
implemented by the application of the same \fb s $\mathbf{H}_{m}, \;m=2,...,M$ to rows and columns of the ``positive" and ``negative" coefficient  arrays. The \cx s from a level $m$ comprise of $4^{m}$ ``positive" blocks of \cx s $\left\{  \mathbf{Z}_{+[m]}^{j,l}\right\},\;l,j=0,...,^{2^{m}-1},$ and the same number of    ``negative" blocks  $\left\{  \mathbf{Z}_{-[m]}^{j,l}\right\}$.

The \cx s from a block are inner products of the signal  $\mathbf{X}=\left\{X[k,n] \right\}\in\Pi[N,N]$ with the shifts of the corresponding \wq :

\begin{equation}\label{inpro}
  \begin{array}{cc}
    Z_{\pm[m]}^{j,l}[k,n]  =&\sum_{\la,\mu=0}^{N-1}X[\la,\mu]\,{\Psi}^{p}_{+\pm[m],j,l}[\la -2^{m}k,\mu -2^{m}n], \\
    Y_{\pm[m]}^{j,l}[k,n] =&\mathfrak{Re}( Z_{\pm[m]}^{j,l}[k,n])= \sum_{\la,\mu=0}^{N-1}X[\la,\mu]\,{\th}^{p}_{\pm[m],j,l}[\la -2^{m}k,\mu -2^{m}n] .
  \end{array}
\end{equation}

The inverse \t s are implemented accordingly. Prior to the \r , some structures, possibly \df t, are defined in the sets  $\left\{\mathbf{Z}_{+[m]}^{j,l}\right\}$ and $\left\{\mathbf{Z}_{-[m]}^{j,l}\right\},\;m=1,...M,$ (for example, 2D \ww,  Best Basis or single-level structures) and some manipulations on the \cx s, (for example, thresholding, shrinkage, $l_1$ minimization) are executed. The \r\  produces two complex arrays
    \(
    \mathbf{X}_{+}\)  and  \(
    \mathbf{X}_{-}\).
The signal  \textbf{X} is restored by $\mathbf{\tilde{X}}=\mathfrak{Re}(\mathbf{X}_{+}+\mathbf{X}_{-})/8$.

\begin{SCfigure}
\caption{Left to right: 1.Image $\mathfrak{Re}(\mathbf{X}_{+})$. 2. Its magnitude DFT \sp um. 3.Image  $\mathfrak{Re}(\mathbf{X}_{-})$. 4.
 Its magnitude DFT \sp um}
  \includegraphics[width=3.5in]{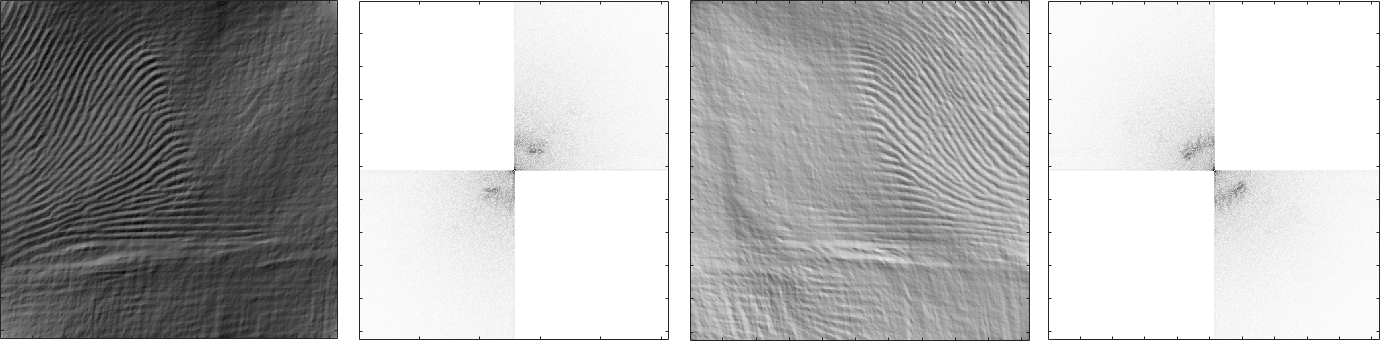}
\label{xp_xm_fi}
\end{SCfigure}

Figure \ref{xp_xm_fi} illustrates the  image ``Fingerprint" restoration by the 2D signals $\mathfrak{Re}(\mathbf{X}_{\pm})$. The signal  $\mathfrak{Re}(\mathbf{X}_{-})$ captures oscillations oriented to \emph{north-east}, while $\mathfrak{Re}(\mathbf{X}_{+})$ captures oscillations oriented to \emph{north-west}. The signal  $\tilde{\mathbf{X}}=\mathfrak{Re}(\mathbf{X}_{+}+\mathbf{X}_{-})/8$ perfectly restores the image achieving  PSNR=312.3538 dB.

\section{Image denoising}\label{sec:s3}
In this section, we describe  application of  the directional qWP \t s  presented in Section \ref{sec:s2} to the
restoration of  an image $\mathbf{X}$  from the data $\mathbf{\check{X}}=\mathbf{X}+\mathbf{E}$, where $\mathbf{E}$ is the Gaussian zero-mean noise whose STD=$\o$.
\subsection{Denoising scheme for 2D qWPs}\label{sec:ss31}
The degraded image $\mathbf{\check{X}}$ is decomposed into two sets $\left\{\mathbf{\check{Z}}_{+[m]}^{j,l}\right\}$ and $\left\{\mathbf{\check{Z}}_{-[m]}^{j,l}\right\},\;m=1,...,M,\;j,l=0,2^{m}-1,$ of the qWP \t\ \cx s, then a version of   the  Bivariate Shrinkage  \al\ (BSA)\cite{bishr,sen_seles} is implemented  and  the image $\tilde{\mathbf{X}}\approx\mathbf{X}$ is restored from the shrunken \cx s.
The restoration is executed separately from the sets of \cx s belonging to several \d\ levels and the results are averaged with some weights.

      \subsubsection{Image restoration from a single-level \t\ \cx s}\label{sec:sss311}
Consider the  \r\ of an image $\mathbf{X}\in\Pi[N,N]$  from  the \underline{fourth}-level \t\ \cx s of the degraded array $\mathbf{\check{X}}$ of size $N\times N$.

     The  \dn\ \al, which we refer to as {qWPdn},  is implemented by the following steps:
    \begin{enumerate}
      \item In order to eliminate boundary effects, the degraded image $\mathbf{\check{X}}$ is \sy ically extended to the image $\mathbf{\check{X}}_{T}$ of size $N_{T}\times N_{T}$, where $N_{T}=N+2T$. Typically, either $T=N/4 \mbox{ or } T=N/8$.
      \item The Bivariate Shrinkage  (BSA) utilizes the interscale dependency of the \t\ \cx s. Therefore, the direct 2D  \t s of the  image $\mathbf{\check{X}}_{T}$ with the complex qWPs $\Psi^{p}_{++}$ and $\Psi^{p}_{+-}$ are executed down to the \underline{fifth} \d\ level.  As a result,   two sets $\mathbf{\check{Z}}_{+[m]}^{j,l}=\left\{\mathbf{\check{Z}}_{+[m]}^{j,l}[k,n]\right\}$ and $\mathbf{\check{Z}}_{-[m]}^{j,l}=\left\{\mathbf{\check{Z}}_{+[m]}^{j,l}[k,n]\right\} ,\;m=1,...,5,\;j,l=0,...,2^{m}-1,\; k,n=0,...,N_{T}/2^{m}-1,$ of the qWP \t\ \cx s are produced.
      \item\label{neval}  The noise variance  is estimated by
        \(\tilde{\o}_{e}^{2}=\frac{\mathrm{median}(|{\check{Z}}_{+[1]}^{1,1}[k,n]|)}{0.6745}. \)
         \item\label{z4+}  Let  $\check{c}_{4}[k,n]\srr\check{Z}_{+[4]}^{j,l}[k,n]$ denote a  coefficient  from a block $\mathbf{\check{Z}}_{+[4]}^{j,l}$ at the fourth \d\ level. The following operations are applied to the coefficient  $\check{c}_{4}[k,n]$:
             \begin{enumerate}
               \item\label{av_var} The averaged variance $\bar{\o}_{c}[k,n]^{2}=\frac{1}{W_{4}^{2}}\sum_{\k,\n=-W_{4}/2}^{W_{4}/2-1}\check{c}_{4}[k+\k,n+\n]^{2}$ is calculated. The integer $W_{4}$ determines the neighborhood of $\check{c}_{4}[k,n]$  size.
               \item The marginal variance for  $\check{c}_{4}[k,n]$ is estimated by $\tilde{\o}[k,n]^{2}=(\bar{\o}_{c}[k,n]^{2}-\tilde{\o}_{e}^{2})_{+}.$\footnote{$s_{+}\srr\max\{s,0\}$.}
               \item In order to estimate  the clean \t\ \cx s from the fourth \d\ level,   \cx s from the fifth level should be  utilized. The size of the coefficient  block $\mathbf{\check{Z}}_{+[4]}^{j,l}$  is $N_{T}/16\times N_{T}/16$. The \cx s from that block are related to the qWP $\Psi^{p}_{++[4],j,l}$, whose \sp um occupies, \as ely, the square $\mathbf{S}_{+[4]}^{j,l}$ of size $N_{T}/32\times N_{T}/32 $ within the quadrant $\mathbf{q}_{0}$ (see Fig. \ref{dia_fipsi}). The \sp um's location determines the directionality of the \we\ $\Psi^{p}_{++[4],j,l}$. On the other hand, four coefficient   blocks $\left\{\mathbf{\check{Z}}_{+[5]}^{2j+\iota,2l+\la}\right\},\;\iota,\la=0,1,$ of size $N_{T}/32\times N_{T}/32 $ are derived by \fr ing the  block $\mathbf{\check{Z}}_{+[4]}^{j,l}$ \cx s followed by downsampling. The \cx s from those blocks are related to the qWPs $\Psi^{p}_{++[5],2j+\iota,2l+\la}$, whose \sp a occupy, \as ely,  the squares $\mathbf{S}_{+[5]}^{2j+\iota,2l+\la}$ of size $N_{T}/64\times N_{T}/64 $, which fill the square $\mathbf{S}_{+[4]}^{j,l}$. Therefore, the orientations of the \we s $\Psi^{p}_{++[5],2j+\iota,2l+\la}$ are close to the orientation of $\Psi^{p}_{++[4],j,l}$. Keeping this  in mind, we form the joint fifth-level array $\mathbf{c}_{5}^{j,l}$ of size $N_{T}/16\times N_{T}/16 $ by interleaving the \cx s from the arrays $\left\{\mathbf{\check{Z}}_{+[5]}^{2j+\iota,2l+\la}\right\}$. To be specific, the joint array $\mathbf{c}_{5}^{j,l}$ consists of the quadruples:
                   \begin{equation*}\label{quad5}
                   \mathbf{c}_{5}^{j,l}=\left\{ \left[
                                                    \begin{array}{cc}
                                                      \check{Z}_{+[5]}^{2j,2l}[\k,\n] &   \check{Z}_{+[5]}^{2j,2l+1}[\k,\n]  \\
                                                        \check{Z}_{+[5]}^{2j+1,2l}[\k,\n]  &   \check{Z}_{+[5]}^{2j+1,2l+1}[\k,\n]  \\
                                                    \end{array}
                                                  \right]
                    \right\},\quad \k,\n=0,...N_{T}/32-1.
                   \end{equation*}
            \item  Let $\breve{c}_{5}[k,n]$ denote a  coefficient  from the joint array $\mathbf{c}_{5}^{j,l}$. Then,         the  \t\ coefficient  $Z^{j,l}_{+[4]}[k,n]$  from the fourth \d\ level is estimated by the bivariate shrinkage of the \cx s $\check{Z}^{j,l}_{+[4]}[k,n]$:
            \begin{equation*}\label{bisr4+}
             Z^{j,l}_{+[4]}[k,n]\approx \tilde{Z}^{j,l}_{+[4]}[k,n]=
             \frac{\left(\sqrt{\check{c}_{4}[k,n]^{2}+\breve{c}_{5}[k,n]^{2}}-
             \frac{\sqrt{3}\,\tilde{\o}_{e}^{2}}{\tilde{\o}[k,n]}\right)_{+}}{\sqrt{\check{c}_{4}[k,n]^{2}+\breve{c}_{5}[k,n]^{2}}}\,\check{c}_{4}[k,n].
            \end{equation*}
            \end{enumerate}
            \item As a result of the above operations, the  fourth-level coefficient  array $\mathbf{\tilde{Z}}_{+[4]}=\left\{\mathbf{\tilde{Z}}_{+[4],j,l}\right\},\;j,l=0,...,15$, is estimated, where $\mathbf{\tilde{Z}}_{+[4],j,l}=\left\{ \tilde{Z}^{j,l}_{+[4]}[k,n] \right\},\;k,n=0,... N_{T}/16-1$.
                \item The inverse qWP \t\ is applied to the coefficient  array $\mathbf{\tilde{Z}}_{+[4]}$ and the result shrinks to the original image size $N\times N$.  Thus, the sub-image $\mathbf{\tilde{X}}_{+}^{4}$ is obtained.
            \item The same operations are applied to $\check{\mathbf{Z}}^{j,l}_{-[m]},\;m=4,5$, thus resulting in the sub-image $\mathbf{\tilde{X}}_{-}^{4}$.

            \item The clean image is estimated by
            \(\mathbf{X}\approx\mathbf{\tilde{X}}^{4}=\frac{\mathfrak{Re}(\mathbf{\tilde{X}}_{+}^{4}+\mathbf{\tilde{X}}_{-}^{4})}{8}.\)
             \end{enumerate}
              \subsubsection{Image restoration from  several \d\ levels}\label{sec:sss312}
              More stable estimation of the image $\mathbf{X}$ is derived by the weighted average of several single-level estimations $\left\{\mathbf{\tilde{X}}^{m}\right\}$. In most cases, the   estimations from the second, third and fourth levels are combined, so that $m=2,3,4$.

              The \as ed image $\mathbf{\tilde{X}}^{3}$ is derived from the third-level \cx s  $\mathbf{\check{Z}}_{\pm[3]}^{j,l}$. The fourth-level \cx s that are needed for the Bivariate Shrinkage of the \cx s $\mathbf{\check{Z}}_{\pm[3]}^{j,l}$ are taken from the ``cleaned" arrays $\mathbf{\tilde{Z}}_{\pm[4],j,l}$ rather than from the ``raw" ones $\mathbf{\check{Z}}_{\pm[4],j,l}$.

              Similarly, the image $\mathbf{\tilde{X}}^{2},$ is derived from the coefficient  arrays   $\mathbf{\check{Z}}_{\pm[2]}^{j,l}$ and $\mathbf{\tilde{Z}}_{\pm[3]}^{j,l}$.
              The final operation is the weighted averaging such as
               \begin{equation}\label{wed3}
             \mathbf{\tilde{X}}=\frac{\a_{2}\mathbf{\tilde{X}}^{2}+\a_{3}\mathbf{\tilde{X}}^{3}+\a_{4}\mathbf{\tilde{X}}^{4}}{\a_{2}+\a_{3}+\a_{4}}.
           \end{equation}

\br\label{rem:time2} Matlab implementation of all the operations needed to \t\ the degraded array $\mathbf{\check{X}}$ of size $512\times512$  into the estimation $ \mathbf{\tilde{X}}$ given by \eh{wed3} takes 1 second. Note that the noise STD is not a part of the input.  It is evaluated as indicated in Item \ref{neval}.\er
\br\label{rem:time3}In some cases,  restoration from third, fourth and fifth levels is preferable. Then, the degraded array $\mathbf{\check{X}}_{T}$  is decomposed down to the sixth level.
\er

\br\label{rem:frepar}The \al\ comprises a number of free parameters which enable a flexible adaptation to the processed class of objects. These parameters are the order ``p" of the generating \s, integers $W_{4}$, $W_{3}$ and $W_{2}$, which determine the sizes of neighborhoods for the  averaged variances calculation, and the weights $\a_{2}$, $\a_{3}$ and $\a_{4}$.\er
   \br\label{rem:ivan} Fragments of the
    Matlab \fd s \texttt{denoising\_dwt.m} and \texttt{bishrink.m} from the websites  http://eeweb.poly.edu/iselesni/WaveletSoftware/denoising\_dwt.html and \\ http://eeweb.poly.edu/iselesni/WaveletSoftware/denoise2.html, respectively, were used as patterns while compiling our own denoising software.\er

\subsection{qWPdn--WNNM: Hybrid \al}\label{sec:ss32}
Equations \rf{inpro} imply that the \cx s of the qWP \t s reflect the correlation of the image under \pr\ with the collection  of waveforms, which are well localized in the spatial domain and are oscillating in multiple \de s with multiple frequencies. By this reason, these \t s are well suited for capturing edges and texture patterns oriented in \df t \de s.
Experiments with the qWPdn  image \dn\ \al\ demonstrate its  ability to restore edges and texture details even from severely degraded images.   In most conducted experiments, the qWPdn provides better resolution of edges and fine structures compared to the most state-of-the-art \al s based on non-local self-similarity (NSS), including the  BM3D, NCSR and WNNM \al s, which is reflected in higher SSIM values. On the other hand, the NSS-based  \al s  proved to be superior in noise suppression especially  in smooth regions in images, thus producing the high PSNR values.

One of the best existing \dn\ \al s is WNNM (\cite{wnnm})  introduced in 2014. It produces high PSNR and SSIM values and, in most cases, a better visual perception of restored images compared to other \al s. WNNM is an iterative image \dn\ \al , whose main components are stacking  non-local  patches that are similar to a given patch  into a low rank matrix, computing the singular value \d\   of this matrix and minimization of the the nuclear norm of the matrix by soft thresholding with adaptive weights the singular values of this matrix.  Even the recently designed methods  (for example, \cite{nlSS}, which exploits the similarity of patches from external images in addition to the inner patches from the given image, and Deep Learning-based methods such as reported in \cite{tnrd, dncnn}) produce marginal, if any, improvements compared to WNNM. Nevertheless, some drawbacks common to NSS-based methods are inherent in WNNM. Namely, some  over-smoothing effect on  the edges and fine texture persists when restoration of  severely degraded images.

We propose to combine  qWPdn with  WNNM \al s to benefit from the strong features of both \al s.

Denote by $\mathbf{Q}$ and $\mathbf{W}$  the operators of application of the qWPdn , which is described in Section \ref{sec:s2}, and WNNM  \dn\ \al s, respectively, to a degraded array $\mathbf{A}$:
$\mathbf{Q}\,\mathbf{A}=\mathbf{D}_{Q}$ and $\mathbf{W}\,\mathbf{A}=\mathbf{D}_{W}$.

Assume that we have an array $\check{\mathbf{X}}^{0}=\mathbf{X}+\mathbf{E}$, which \ry s an image $\mathbf{X}$ degraded by additive Gaussian noise $\mathbf{E}$ whose STD is $\o.$
The \dn\ processing is implemented along  the following cross-boosting scheme.

\begin{description}
  \item[First step:]  Apply the  operators $\mathbf{Q}$ and $\mathbf{W}$ to the input array $\check{\mathbf{X}}^{0}$: $\mathbf{Y}_{Q}^{1}=\mathbf{Q}\,\check{\mathbf{X}}^{0}$ and $\mathbf{Y}_{W}^{1}=\mathbf{W}\,\check{\mathbf{X}}^{0}$.
  \item[Iterations:] $i=1,...,I-1$
 \begin{enumerate}
  \item Form new  input arrays
  \(\check{\mathbf{X}}_{Q}^{i}=\frac{\check{\mathbf{X}}^{0}+\mathbf{Y}_{Q}^{i}}{2},\quad \check{\mathbf{X}}_{W}^{i}=\frac{\check{\mathbf{X}}^{0}+\mathbf{Y}_{W}^{i}}{2}.\)
  \item Apply the  operators $\mathbf{Q}$ and $\mathbf{W}$ to the  input arrays:
  \(\mathbf{Y}_{Q}^{i+1}=\mathbf{Q}\,\check{\mathbf{X}}_{W}^{i}, \quad \mathbf{Y}_{W}^{i+1}=\mathbf{W}\,\check{\mathbf{X}}_{Q}^{i}. \)
\end{enumerate}
  \item[Estimations of the clean image:] Three estimations are used:
  \begin{enumerate}
    \item The cross-boosted WNNM  estimation $\tilde{\mathbf{X}}_{uW}\srr \mathbf{Y}_{W}^{I}$ (\textbf{cbWNNM }).
    \item The cross-boosted qWPdn estimation $\tilde{\mathbf{X}}_{uQ}\srr \mathbf{Y}_{Q}^{I}$ (\textbf{cbqWP}).
    \item The hybrid estimation $\tilde{\mathbf{X}}_{H}\srr (\mathbf{Y}_{W}^{I}+\mathbf{Y}_{Q}^{I})/2$   (\textbf{hybrid}).
  \end{enumerate}
\end{description}

\subsection{Experimental results}\label{sec:ss33}
In this section, we compare the performance of our \dn\ schemes designated as  \textbf{cbWNNM }, \textbf{cbqWP} and \textbf{hybrid} on the restoration of degraded images with the performances of  the state-of-the-art \al s such as  {BM3D} (\cite{bm3d}), {BM3D-SAPCA} (\cite{sapca}), {WNNM} (\cite{wnnm}),  {NCSR} (\cite{ncsr}), cptTP-$\mathbb{C}$TF$_{6}$ (\cite{Zhu_han}) and {DAS-2}  (\cite{che_zhuang}).
To produce results for the comparison, we used the software available at the websites  http://www.cs.tut.fi/~foi/GCF-BM3D/index.html\#ref\_software (BM3D and BM3D-SAPCA),  http://staffweb1.cityu.edu.hk/xzhuang7/softs/index.html\#bdTPCTF  (cptTP-$\mathbb{C}TF_{6}$ and DAS-2), https://github.com/csjunxu/WNNM\_CVPR2014 (WNNM),  and \\ https://www4.comp.polyu.edu.hk/~cslzhang/NCSR.htm (NCSR).

 The restored images were evaluated by the visual perception, by
 Peak Signal-to-Noise ratio (PSNR) (see \eh{psnr})\footnote{\begin{equation}\label{psnr}
 PSNR(\mathbf{x},\mathbf{\tilde{x}})\srr10\log_{10}\left(\frac{K\,255^2}{\sum_{k=1}^K(x_{k}-\tilde
  x_{k})^2}\right)\; dB.
  \end{equation}}  and by the  Structural Similarity  Index (SSIM) (\cite{ssim}, it is computed by the \texttt{ssim.m} Matlab 2020b \fd). The SSIM measures the structural similarity of small moving windows in two images. It varies from 1 for fully identical windows to -1 for completely dissimilar ones. The  global index is the average of  local indices. Currently,  SSIM is regarded as more informative characteristics of the image quality compared to PSNR  and Mean Square Error (MSE) (see discussion in \cite{PS_MS}).

\begin{figure}
\centering
\includegraphics[width=6.5in]{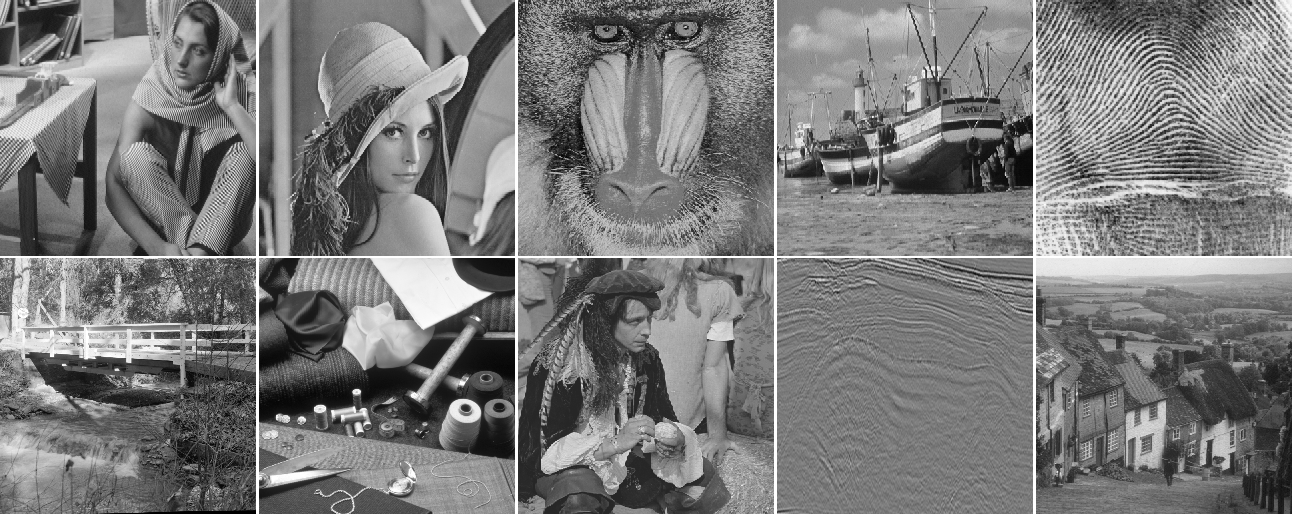}
\caption{Clean images: ``Lena",   ``Boat",   ``Hill, ``Barbara",  ``Mandrill",  ``Bridge", ``Man", ``Fabric",  ``Fingerprint" and   ``Seismic"}
\label{clima}
\end{figure}

For the experiments, we used a standard set of benchmark images: ``Lena",   ``Boat",     ``Hill", ``Barbara", ``Mandrill",  ``Bridge", ``Man", ``Fabric"  and
 ``Fingerprint".
 One image that represents a stacked seismic section is  designated as ``Seismic".
 The ''clean" images are displayed in Fig. \ref{clima}.

\begin{figure}
\centering
\includegraphics[width=6.0in]{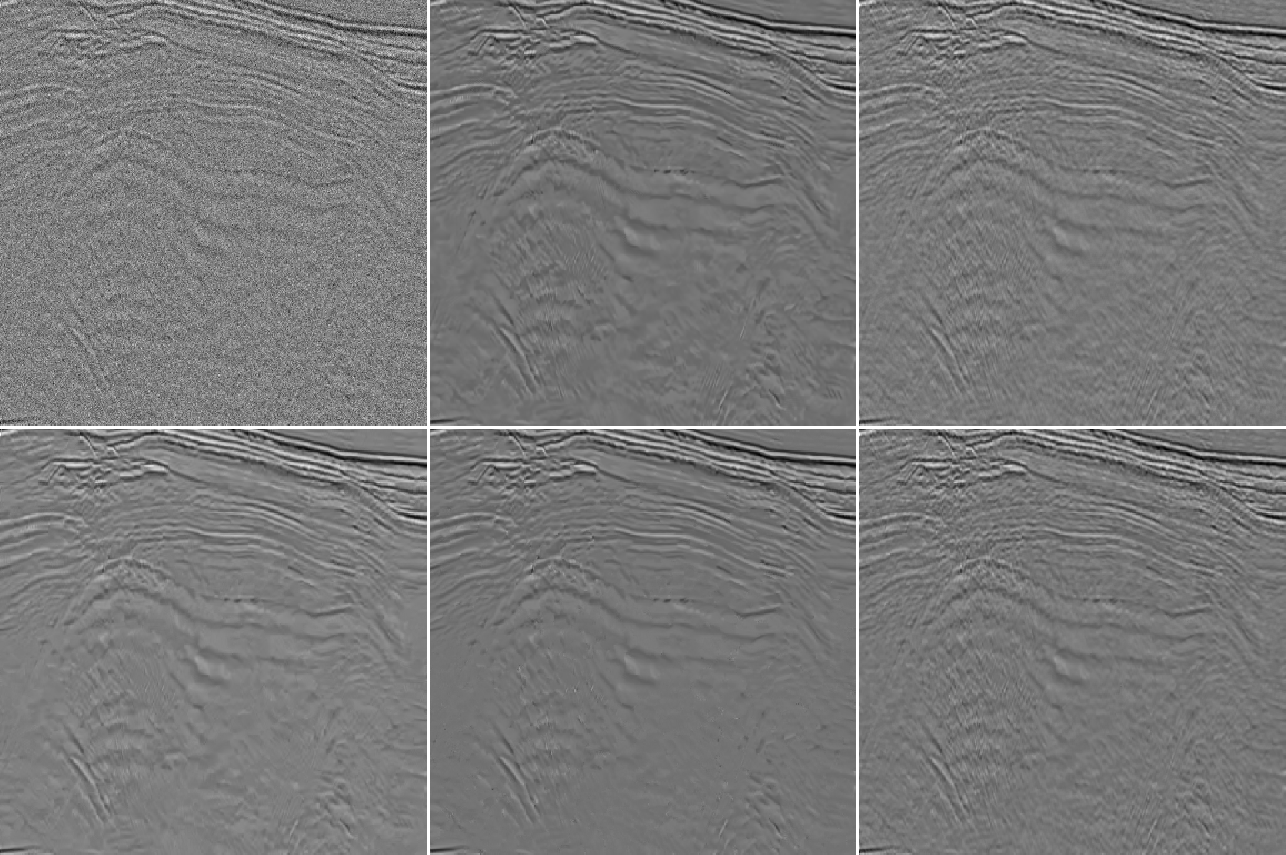}
\includegraphics[width=6.0in]{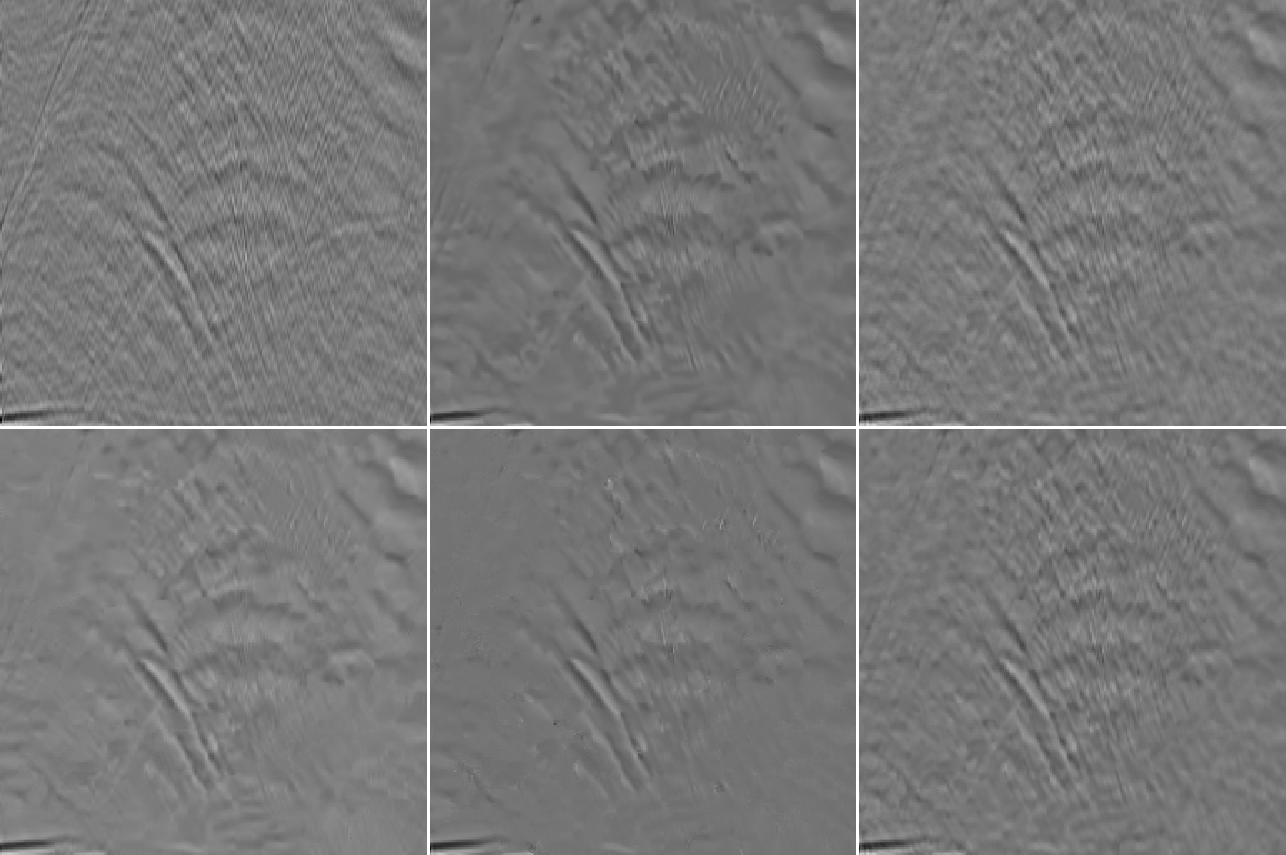}
\caption{Restoration of ``Seismic" image corrupted by Gaussian noise with STD $\o=25$ dB. PSNR/SSIM for {BM3D}-- 30.8/0.4984; for {WNNM}-- 30.65/0.4556; for {BM3D-SAPCA}--30.8/0.5016;  for \textbf{cbqWP}--30.89/{0.5849}; for  \textbf{hybrid}--{30.99}/0.5745}
\label{seis25_6}
\end{figure}

The images were corrupted by  Gaussian zero-mean noise whose STD was $\o=$5, 10, 25, 40, 50, 80 and 100 dB. Then, the  {BM3D}, {BM3D-SAPCA}, {NCSR}, {WNNM}, {cbWNNM },  {cbqWP},  {hybrid},   cptTP-$\mathbb{C}TF_{6}$ and{ DAS-2} \dn\ \al s were applied to restore the images. In most experiments  the \al\ {cbWNNM } performed better than   {cbqWP}. However, this was not the case in experiments with the ``Seismic" image. Therefore, in the  ``Seismic" block in Table \ref{tabman} and pictures in Fig.   \ref{seis25_6} we provide results from experiments with the {cbqWP} rather than with the  {cbWNNM } \al.

Table \ref{tabbar} summarizes  experimental results from the restoration of the ``Barbara",  ``Boat", ``Fingerprint", ``Lena"  and ``Mandrill"  images  corrupted by additive Gaussian noise. PSNR and SSIM values for each experiment are given.

\begin{table}

\caption{{PSNR}/{SSIM} values from restoration of ``Barbara",  ``Boat", ``Fingerprint", ``Lena"  and ``Mandrill"  images. Boldface highlights  the best results. Noise STD=$\sigma$}\label{tabbar}
\resizebox{\columnwidth}{!}{
  \centering
  \begin{tabular}{|c|c|c|c|c|c|c|c|}
    \hline
$\sigma$ & 5 & 10 & 25 & 40 & 50 & 80 & 100\\\hline
\multicolumn{8}{|c|}{Barbara}\\
 \hline
noised &${34.16}/{0.77}$ & ${28.14}/{0.6}$  & ${20.18}/{0.35}$  & ${16.09}/{0.23}$  & ${14.16}/{0.18}$  & ${10.07}/{0.09}$  & ${8.14}/{0.06}$ \\\hline
\vspace{0.1cm}
WNNM &$\mathbf{{38.77}}/{0.836}$ & $\mathbf{35.51}/{0.7733}$  & ${\mathbf{31.27}}/{0.68}$  & ${{28.77}/{0.5993}}$  & ${{27.8}}/{0.5628}$  & ${{25.48}/{0.4499}}$  & ${24.36}/{0.3891}$ \\\hline\vspace{0.1cm}
SAPCA &${38.37}/{0.8445}$ & ${35.1}/{0.7756}$  & ${{31}}/{0.6797}$  & ${{28.68}/{0.6086}}$  & ${{27.5}}/{0.5605}$  & ${24.3}/{0.402}$  & ${23.09}/{0.339}$ \\\hline\vspace{0.1cm}
NCSR &${38.37}/{0.8302}$ & ${35.4}/{0.7674}$  & ${30.62}/{0.6579}$  & ${28.19}/{0.5687}$  & ${27}/{0.5176}$  & ${24.37}/{0.3902}$  & ${23.21}/{0.3296}$ \\\hline\vspace{0.1cm}
cptTP-$\mathbb{C}$TF$_6$ &${37.71}/{0.8382}$ & ${34.04}/{0.763}$  & ${29.24}/{0.6358}$  & ${26.75}/{0.5375}$  & ${25.61}/{0.4833}$  & ${23.44}/{0.3633}$  & ${22.55}/{0.3092}$ \\\hline\vspace{0.1cm}
DAS-2&${37.7}/{0.8438}$ & ${33.98}/{0.764}$  & ${29.4}/{0.6422}$  & ${27.09}/{0.5595}$  & ${26.01}/{0.5121}$  & ${23.72}/{0.3987}$  & ${22.63}/{0.3398}$ \\\hline
\vspace{0.1cm}
BM3D&${38.3}/{0.8387}$ & ${34.97}/{0.7738}$  & ${{30.71}}/{0.6707}$  & ${27.98}/{0.5788}$  & ${27.23}/{0.5389}$  & ${24.79}/{0.4192}$  & ${23.63}/{0.3608}$ \\\hline
\vspace{0.1cm}
cbWNNM&${38.74}/{{0.853}}$ & ${{35.4}}/{0.7833}$  & ${31.11}/{\mathbf{0.6852}}$  & $\mathbf{{28.9}/{0.6171}}$  & $\mathbf{27.81}/{\mathbf{0.576}}$  & ${\mathbf{25.51}/{0.4681}}$  & $\mathbf{24.47}/{{0.4076}}$ \\\hline
\vspace{0.1cm}
Hybrid&${{38.59}}/\mathbf{0.8543}$ & ${35.25}/{\mathbf{0.7859}}$  & ${30.87}/{0.6834}$  & ${28.67}/{0.6154}$  & ${27.59}/{{0.5745}}$  & ${{25.35}/\mathbf{0.4688}}$  & ${24.33}/{{\mathbf{ 0.4114}}}$ \\\hline 
  \multicolumn{8}{|c|}{Boat}\\
    \hline
    noised &${34.16}/{0.79}$ & ${28.14}/{0.58}$  & ${20.18}/{0.29}$  & ${16.09}/{0.18}$  & ${14.16}/{0.13}$  & ${10.07}/{0.07}$  & ${8.14}/{0.05}$ \\\hline
\vspace{0.1cm}
WNNM&${{37.35}}/{0.8042}$ & ${{34.07}}/{0.6803}$  & ${\mathbf{30.05}}/{0.5266}$  & ${\mathbf{27.96}}/{0.4317}$  & ${{26.98}}/{0.3833}$  & ${25.03}/{0.2853}$  & ${24.11}/{0.2389}$ \\\hline\vspace{0.1cm}
SAPCA&${\mathbf{37.51}}/\mathbf{0.8288}$ & ${\mathbf{34.1}}/{0.6925}$  & ${{30.03}}/{0.5282}$  & ${{27.92}}/{0.4384}$  & ${{26.89}}/{0.3931}$  & ${24.68}/{0.2993}$  & ${23.69}/{0.2609}$ \\\hline\vspace{0.1cm}
NCSR&${37.33}/{0.8101}$ & ${33.92}/{0.6848}$  & ${29.78}/{0.5053}$  & ${27.64}/{0.4003}$  & ${ 26.4}/{0.3551}$  & ${24.61}/{0.257}$  & ${23.69}/{0.217}$ \\\hline\vspace{0.1cm}
cptTP-$\mathbb{C}$TF$_6$ &${36.86}/{0.8044}$ & ${33.34}/{0.6713}$  & ${29.12}/{0.504}$  & ${27.02}/{0.4044}$  & ${26.09}/{0.3567}$  & ${24.27}/{0.2626}$  & ${23.48}/{0.2232}$ \\\hline\vspace{0.1cm}
DAS-2& ${36.85}/{0.8294}$  & ${33.15}/{0.688}$  & ${28.88}/{0.5182}$  & ${26.78}/{0.4228}$  & ${25.81}/{0.3763}$
 & ${23.84}/{0.2792}$&${22.9}/{0.2361}$
\\\hline
\vspace{0.1cm}
BM3D&${{37.29}}/{0.8065}$ & ${{33.91}}/{0.6805}$  & ${{29.9}}/{0.5296}$  & ${27.74}/{0.4395}$  & ${26.77}/{0.3899}$  & ${{24.87}}/{0.2952}$  & ${23.96}/{0.2533}$ \\\hline
\vspace{0.1cm}
cbWNNM&${37.42}/{{0.8257}}$ & ${34.03}/{0.702}$  & ${29.99}/{0.5455}$  & ${{27.95}}/{0.4568}$  & ${\mathbf{27}}/{0.4113}$  & ${\mathbf{25.09}}/{0.3116}$  & $\mathbf{24.26}/{{0.2638}}$ \\\hline
\vspace{0.1cm}
Hybrid&${{37.13}/{0.8194}}$ & ${33.86}/{\mathbf{0.7105}}$  & ${29.72}/{\mathbf{0.5521}}$  & ${27.66}/{\mathbf{0.4616}}$  & ${26.78}/{\mathbf{0.4158}}$  & ${24.92}/{\mathbf{0.3182}}$  & ${24.14}/\mathbf{0.2725}$ \\\hline
  \multicolumn{8}{|c|}{Fingerprint}\\
    \hline
noised &${34.16}/{0.97}$ & ${28.14}/{0.91}$  & ${20.18}/{0.67}$  & ${16.09}/{0.48}$  & ${14.16}/{0.38}$  & ${10.07}/{0.21}$  & ${8.14}/{0.15}$ \\\hline
\vspace{0.1cm}
WNNM  &${{35.16}/{0.9801}}$ & ${{32.68}/\mathbf{0.9655}}$  & $\mathbf{27.94}/{0.9014}$  & ${25.6}/{0.8388}$  & $\mathbf{24.7}/{0.8093}$  & ${22.77}/{0.7289}$  & ${21.85}/{0.6777}$ \\\hline\vspace{0.1cm}
SAPCA  &${36.66}/{0.9859}$ & ${32.64}/{0.9648}$  & ${27.8}/{0.8981}$  & ${25.54}/{0.839}$  & ${24.53}/{0.8057}$  & ${22.3}/{0.7068}$  & ${21.18}/{0.6449}$ \\\hline\vspace{0.1cm}
NCSR  &${{36.8}}/{0.9860}$ & ${\mathbf{32.69}}/{0.9645}$  & ${{27.83}}/{0.8968}$  & ${25.52}/{0.8290}$  & ${24.48}/{0.7894}$  & ${22.37}/{0.6869}$  & ${21.4}/{0.6295}$ \\\hline\vspace{0.1cm}
cptTP-$\mathbb{C}$TF$_6$ &${36}/{0.9845}$ & ${32.23}/{0.9619}$  & ${27.33}/{0.889}$  & ${25.07}/{0.8219}$  & ${24.04}/{0.7815}$  & ${21.98}/{0.6746}$  & ${21.03}/{0.6129}$ \\\hline\vspace{0.1cm}
DAS-2&${36.25}/{0.9843}$ & ${32.03}/{0.9601}$  & ${27.07}/{0.886}$  & ${24.86}/{0.8246}$  & ${23.88}/{0.7895}$  & ${21.87}/{0.7013}$ & ${20.96}/{0.652}$
\\\hline
\vspace{0.1cm}
BM3D&${36.5}/{0.9854}$ & ${32.45}/{0.9634}$  & ${27.71}/{0.8955}$  & ${25.3}/{0.8334}$  & ${24.53}/{0.8019}$  & ${22.2}/{0.7165}$  & ${21.61}/{0.6643}$ \\\hline
\vspace{0.1cm}
cbWNNM&$\mathbf{36.91}/\mathbf{0.9864}$ & ${32.1}/{0.9615}$  & ${27.85}/{0.9039}$ & ${\mathbf{25.68}/{0.851}}$
 & ${{24.67}}/\mathbf{0.8215}$  & ${\mathbf{22.79}/{0.7417}}$  & ${\mathbf{21.89}/{{0.7031}}}$ \\\hline
\vspace{0.1cm}
Hybrid&${36.8}/{{0.9861}}$ & ${32.51}/{{0.9645}}$  & ${27.88}/{\mathbf{0.904}}$  & ${25.63}/\mathbf{{0.8512}}$
 & ${24.59}/{{0.8213}}$  & ${22.73}/\mathbf{{0.7425}}$  & ${21.77}/\mathbf{0.7061}$ \\\hline
 \multicolumn{8}{|c|}{Lena}\\
    \hline
noised &${34.16}/{0.65}$ & ${28.14}/{0.43}$  & ${20.18}/{0.2}$  & ${16.09}/{0.12}$  & ${14.16}/{0.09}$  & ${10.07}/{0.04}$  & ${8.14}/{0.03}$ \\\hline
\vspace{0.1cm}
WNNM &${\mathbf{38.8}}/{0.7131}$ & ${{36.05}}/{0.6193}$  & ${\mathbf{32.25}}/{0.5036}$  & ${{30.1}}/{0.425}$  & ${{29.24}}/{0.3942}$  & ${27.22}/{0.3208}$  & ${26.19}/{0.278}$ \\\hline\vspace{0.1cm}
SAPCA&${{38.67}}/{0.6893}$ & ${\mathbf{36.07}}/{0.6233}$  & ${{32.23}}/{0.5077}$  & ${{30.11}}/{0.4411}$  & ${{29.07}}/{0.408}$  & ${26.5}/{0.314}$  & ${25.1}/{0.2771}$ \\\hline\vspace{0.1cm}
NCSR&${38.73}/{0.7123}$ & ${ 35.85}/{0.6214}$  & ${ 31.92 }/{0.4841}$  & ${ 29.91}/{0.4155}$  & ${28.9}/{0.382}$  & ${26.72}/{0.3025}$  & ${25.71}/{0.2685}$ \\\hline\vspace{0.1cm}
cptTP-$\mathbb{C}$TF$_6$ &${38}/{0.705}$ & ${35.46}/{0.616}$  & ${31.53}/{0.4992}$  & ${29.41}/{0.4313}$  & ${28.41}/{0.3965}$  & ${26.36}/{0.3199}$  & ${25.41}/{0.2833}$ \\\hline\vspace{0.1cm}
DAS-2& ${38.18}/{0.7272}$  & ${35.2}/{0.625}$  & ${31.09}/{0.4920}$  & ${28.9}/{0.4148}$  & ${27.84}/{0.3757}$
 & ${25.53}/{0.29}$&${24.4}/{0.2285}$
\\\hline
\vspace{0.1cm}
BM3D&${{38.72}}/{0.7078}$ & ${{35.92}}/{0.6233}$  & ${{32.07}}/{0.5026}$  & ${29.87}/{0.4265}$  & ${{29.04}}/{0.3957}$  & ${{26.98}}/{0.3214}$  & ${{25.95}}/{0.2851}$  \\\hline
\vspace{0.1cm}
cbWNNM&${38.79}/{{0.7095}}$ & ${36}/{0.633}$  & ${32.20}/{{0.5155}}$  & ${\mathbf{30.2}}/{0.4479}$  & ${\mathbf{29.25}/{0.4139}}$  & ${\mathbf{27.24}/\mathbf{0.3359}}$  & ${\mathbf{26.36}/{0.304}}$ \\\hline
\vspace{0.1cm}
Hybrid&${{38.76}}/\mathbf{0.7218}$ & ${35.89}/{\mathbf{0.6408}}$  & ${32.02}/{\mathbf{0.5202}}$  & ${30}/{\mathbf{0.4528}}$  & ${29.06}/{\mathbf{0.4193}}$  & ${27.07}/{{0.3349}}$  & ${26.21}/{\mathbf{0.312}}$ \\\hline
 \multicolumn{8}{|c|}{Mandrill}\\
    \hline
noised &${34.16}/{0.7}$ & ${28.14}/{0.52}$  & ${20.18}/{0.51}$  & ${16.09}/{0.34}$  & ${14.16}/{0.26}$  & ${10.07}/{0.14}$  & ${8.14}/{0.1}$ \\\hline
\vspace{0.1cm}
WNNM  &${{34.7}}/{0.8902}$ & ${{30.36}}/{0.7994}$  & ${{25.44}}/{0.605}$  & ${\mathbf{23.43}}/{0.4739}$  & ${\mathbf{22.58}}/{0.4024}$  & ${\mathbf{21.1}}/{0.2645}$  & $\mathbf{20.46}/{0.1949}$
\\\hline\vspace{0.1cm}
SAPCA &${\mathbf{35.2}}/{0.9281}$ & ${\mathbf{30.59}}/{0.8278}$  & ${\mathbf{25.54}}/{0.6244}$  & ${{23.4}}/{0.4843}$  & ${{22.54}}/{0.413}$  & ${{20.92}}/{0.2649}$  & ${20.3}/{0.2048}$
\\\hline\vspace{0.1cm}
NCSR&${35.07}/{0.9129}$ & ${30.38}/{0.7986}$  & ${25.36}/{0.5908}$  & ${23.18}/{0.4325}$  & ${22.35}/{0.3669}$  & ${20.82}/{0.2215}$  & ${20.23}/{0.162}$
\\\hline\vspace{0.1cm}
cptTP-$\mathbb{C}$TF$_6$ &${35.06}/{0.9252}$ & ${30.32}/{0.8198}$  & ${25.3}/{0.6082}$  & ${23.16}/{0.4531}$  & ${22.25}/{0.371}$  & ${20.72}/{0.2148}$  & ${20.2}/{0.1617}$ \\\hline\vspace{0.1cm}
DAS-2& ${35.02}/{\mathbf{0.9301}}$  & ${30.24}/{{0.8329}}$  & ${25.24}/{0.64}$  & ${23.2}/{0.51}$  & ${22.34}/{0.4423}$
 & ${20.8}/{0.2997}$&${20.14}/{0.2376}$
\\\hline
\vspace{0.1cm}
BM3D&${{{34.98}}}/{0.9209}$ & ${{30.34}}/{0.8135}$  & ${{25.27}}/{0.6095}$  & ${23}/{0.4613}$  & ${{22.27}}/{0.3813}$  & ${{20.9}}/{0.2439}$  & ${{20.39}}/{0.1956}$ \\\hline
\vspace{0.1cm}
cbWNNM&${34.34}/{{0.9158}}$ & ${{30.39}}/{\mathbf{0.8359}}$  & ${25.53}/{\mathbf{0.6631}}$  & ${{23.35}}/\mathbf{0.5477}$  & ${{{22.39}}/{0.4802}}$  & ${{{20.68}}/{0.3513}}$  & ${{20.03}/{0.2841}}$ \\\hline
\vspace{0.1cm}
Hybrid&${{33.23}}/{0.9028}$ & ${30.02}/{{0.8303}}$  & ${25.4}/{{0.6625}}$  & ${23.14}/{{0.5471}}$  & ${22.28}/{\mathbf{0.4887}}$  & ${20.31}/{\mathbf{0.3565}}$  & ${19.59}/{\mathbf{0.2945}}$ \\\hline
\end{tabular}
}
\end{table}

Table \ref{tabman} summarizes  experimental results from the restoration of the ``Hill",  ``Seismic",  ``Fabric",  ``Bridge"   and``Man"   images corrupted by additive Gaussian noise.  The PSNR and SSIM values for each experiment are given.

\begin{table}
\caption{{PSNR}/{SSIM} values from restoration of ``Hill",  ``Seismic",  ``Fabric",  ``Bridge"   and``Man"  images. Boldface highlights  the best results. Noise STD=$\sigma$.}\label{tabman}
\resizebox{\columnwidth}{!}{
  \centering
  \begin{tabular}{|c|c|c|c|c|c|c|c|}
    \hline
$\sigma $& 5 & 10 & 25 & 40 & 50 & 80 & 100\\\hline
\multicolumn{8}{|c|}{Hill}\\
 \hline
 noised &${34.16}/{0.8}$ & ${28.14}/{0.59}$  & ${20.18}/{0.27}$  & ${16.09}/{0.15}$  & ${14.16}/{0.11}$  & ${10.07}/{0.05}$  & ${8.14}/{0.03}$ \\\hline
\vspace{0.1cm}
WNNM&${{37.02}}/{0.8239}$ & ${{33.7}}/{0.7121}$  & ${{29.96}}/{0.5225}$  & ${\mathbf{28.18}}/{0.4203}$  & ${27.34}/{0.373}$  & ${25.63}/{0.2822}$  & ${24.75}/{0.236}$
\\\hline\vspace{0.1cm}
SAPCA&${\mathbf{37.3}}/{0.8457}$ & ${\mathbf{33.84}}/{0.7316}$  & ${{29.95}}/{0.5335}$  & ${{28.09}}/{0.4281}$  & ${27.2}/{0.3815}$  & ${25.18}/{0.2867}$  & ${24.29}/{0.2491}$
\\\hline\vspace{0.1cm}
NCSR&${37.17}/{0.8350}$ & ${33.7}/{0.7246}$  & ${29.72}/{0.5138}$  & ${27.84}/{0.3936}$  & ${ 26.99}/{0.3463}$  & ${25.16}/{0.2480}$  & ${24.35}/{0.2089}$
\\\hline\vspace{0.1cm}
cptTP-$\mathbb{C}$TF$_6$ &${36.83}/{0.8388}$ & ${33.17}/{0.7157}$  & ${29.13}/{0.4992}$  & ${27.37}/{0.3903}$  & ${26.59}/{0.3425}$  & ${25}/{0.2495}$  & ${24.29}/{0.2108}$
\\\hline\vspace{0.1cm}
DAS-2& ${36.68}/{{0.8447}}$  & ${33.02}/{0.7292}$  & ${29.15}/{0.5327}$  & ${27.31}/{0.4245}$  & ${26.47}/{0.3747}$
 & ${24.62}/{0.2733}$&${23.7}/{0.2272}$
\\\hline
\vspace{0.1cm}
BM3D&${37.14}/{0.8384}$ & ${{33.61}}/{0.7193}$  & ${{29.85}}/{0.5311}$  & ${27.99}/{0.4305}$  & ${{27.19}}/{0.3825}$  & ${{25.42}}/{0.2878}$  & ${{24.59}}/{0.2448}$ \\\hline
\vspace{0.1cm}
cbWNNM&$37.2/{{0.8515}}$ & ${33.72}/0.7461$  & $\mathbf{30.01}/{0.5545}$ & ${{28.13}}/{0.4584}$ & ${{\mathbf{27.4}}/{0.4033}}$  & ${\mathbf{25.64}}/{0.3148}$  & ${{\mathbf{24.87}}/{0.2718}}$ \\\hline
\vspace{0.1cm}
Hybrid&${{37.09}}/\mathbf{0.8516}$ & ${33.64}/{\mathbf{0.7511}}$  & ${29.86}/{\mathbf{0.5655}}$  & ${28.01}/{\mathbf{0.4642}}$  & ${27.29}/{\mathbf{0.4118}}$  & ${{25.62}}/{\mathbf{0.3215}}$  & ${24.84}/{\mathbf{0.278}}$ \\\hline
\hline
  \multicolumn{8}{|c|}{Seismic}\\
    \hline
noised &${34.16}/{0.8}$ & ${28.14}/{0.55}$  & ${20.18}/{0.22}$  & ${16.09}/{0.12}$  & ${14.16}/{0.08}$  & ${10.07}/{0.04}$  & ${8.14}/{0.03}$ \\\hline
\vspace{0.1cm}
WNNM&${39.26}/{0.9205}$ & ${34.95}/{0.765}$  & ${30.65}/{0.4556}$  & ${28.93}/{0.3092}$  & ${28.34}/{0.2524}$  & $\mathbf{27.11}/{0.1621}$  & $\mathbf{26.56}/{0.1305}$
\\\hline\vspace{0.1cm}
SAPCA&${39.09}/{0.9167}$ & ${35.04}/{0.7871}$  & ${30.8}/{0.5016}$  & ${29.04}/{0.3654}$  & ${28.2}/{0.3124}$  & ${26.34}/{0.23}$  & ${25.4}/{0.1934}$
\\\hline\vspace{0.1cm}
NCSR&${38.97}/{0.9131}$ & ${34.86}/{0.78}$  & ${30.62}/{0.4643}$  & ${29}/{0.2948}$  & ${28.27}/{0.2374}$  & ${26.81}/{0.139}$  & ${26.25}/{0.1089}$
\\\hline\vspace{0.1cm}
cptTP-$\mathbb{C}$TF$_6$ &${38.81}/{0.9107}$ & ${34.83}/{0.7696}$  & ${30.72}/{0.4616}$  & ${29.1}/{0.3188}$  & ${28.4}/{0.2618}$  & ${26.91}/{0.1616}$  & ${26.16}/{0.127}$
\\\hline\vspace{0.1cm}
DAS-2& ${38.8}/{0.9141}$  & ${34.83}/{0.795}$  & ${30.67}/{0.5383}$  & ${28.87}/{0.3938}$  & ${28.05}/{0.3318}$
 & ${26.22}/{0.2222}$&${25.29}/{0.1785}$
\\\hline
\vspace{0.1cm}
BM3D&${39}/{0.9126}$ & ${{34.9}}/{0.77}$  & ${{30.8}}/{0.4984}$  & ${29.08}/{0.3606}$  & ${\mathbf{28.45}}/{0.2853}$  & ${{26.97}}/{0.19}$  & ${{26.31}}/{0.1547}$ \\\hline
\vspace{0.1cm}
cbqWP&$\mathbf{{39.57}/{0.9318}}$ & $\mathbf{{35.41}/{0.8321}}$  & ${30.89}/{\mathbf{0.5849}}$  & ${{28.98}}/{\mathbf{0.445}}$  & ${{27.97}}/{\mathbf{0.3836}}$  & ${25.42}/{\mathbf{0.2755}}$
 & ${23.36}/{\mathbf{0.2299}}$ \\\hline
\vspace{0.1cm}
Hybrid&${{{39.13}}/{0.9241}}$ & ${{35.41}/{0.8266}}$  & ${\mathbf{30.99}}/{{0.5745}}$
 & ${\mathbf{29.26}}/{{0.4298}}$  & ${{28.4}}/{{0.369}}$  & ${26.28}/{{0.2657}}$  & ${24.75}/{{{0.2227}}}$ \\\hline

  \multicolumn{8}{|c|}{Fabric}\\\hline
noised &${34.16}/{0.8}$ & ${28.14}/{0.55}$  & ${20.18}/{0.22}$  & ${16.09}/{0.12}$  & ${14.16}/{0.08}$  & ${10.07}/{0.04}$  & ${8.14}/{0.03}$ \\\hline
\vspace{0.1cm}
WNNM&${\mathbf{38.75}/{0.833}}$ & ${{34.92}/{0.7318}}$  & ${{30.76}/{0.5488}}$  & ${{28.83}/{0.4485}}$  & ${\mathbf{28.01}}/{0.4113}$  & $\mathbf{26.19}/{0.3361}$  & $\mathbf{25.26}/{0.2992}$
\\\hline\vspace{0.1cm}
SAPCA&${38.72}/{0.8351}$ & $\mathbf{{35}/{0.7432}}$  & $\mathbf{{30.8}/{0.5678}}$  & ${{28.82}/{\mathbf{0.4726}}}$  & ${{27.89}}/{0.3905}$  & ${25.54}/{0.3275}$  & ${24.51}/{0.2915}$
\\\hline\vspace{0.1cm}
NCSR&${38.58}/{0.8263}$ & ${34.78}/{0.7298}$  & ${30.53}/{0.5408}$  & ${28.54}/{0.4384}$  & ${27.69}/{0.3988}$  & ${25.77}/{0.3185}$  & ${24.8}/{0.2838}$
\\\hline\vspace{0.1cm}
cptTP-$\mathbb{C}$TF$_6$ &${37.99}/{0.81}$ & ${34.07}/{0.6978}$  & ${29.73}/{0.5127}$  & ${27.83}/{0.4266}$  & ${26.94}/{0.3905}$  & ${25.12}/{0.3185}$  & ${24.23}/{0.2856}$
\\\hline\vspace{0.1cm}
DAS-2& ${37.74}/{0.8017}$  & ${33.78}/{0.6985}$  & ${29.48}/{0.5223}$  & ${27.48}/{0.4283}$  & ${26.52}/{0.3858}$
 & ${24.45}/{0.3017}$&${23.42}/{0.2641}$
\\\hline
\vspace{0.1cm}
BM3D&${38.47}/{0.8303}$ & ${{34.63}}/{0.7262}$  & ${{30.57}}/{0.5476}$  & ${{28.65}}/{0.4548}$  & ${{27.85}}/{0.4182}$  & ${{26}}/{0.3402}$  & ${{25.06}}/{0.3032}$ \\\hline
\vspace{0.1cm}
cbWNNM&${{38.69}/\mathbf{0.8354}}$ & ${{34.92}/{0.7426}}$  & ${30.74}/{{0.5649}}$  & ${\mathbf{28.84}}/{\mathbf{0.4726}}$  & ${27.96}/{{0.4279}}$  & ${26.15}/{{0.3489}}$  & $\mathbf{25.26}/{\mathbf{0.3128}}$ \\\hline
\vspace{0.1cm}
Hybrid&${{38.52}}/{0.8294}$ & ${34.76}/{{0.7349}}$  & ${{30.52}}/{0.561}$  & ${{28.61}}/{0.4718}$  & ${{27.73}}/\mathbf{0.4283}$  & ${25.9}/{\mathbf{0.352}}$  & ${25.01}/{{0.3096}}$
\\\hline
  \multicolumn{8}{|c|}{Bridge}\\
    \hline
noised &${34.16}/{0.7}$ & ${28.14}/{0.52}$  & ${20.18}/{0.51}$  & ${16.09}/{0.34}$  & ${14.16}/{0.26}$  & ${10.07}/{0.14}$  & ${8.14}/{0.1}$ \\\hline
\vspace{0.1cm}
WNNM&${35.81}/{0.9387}$ & ${31.17}/{0.854}$  & ${26.29}/{0.635}$  & ${{24.45}}/{0.4992}$  & ${23.67}/{0.43}$  & $\mathbf{22.24}/{0.3095}$  & ${21.6}/{0.2568}$
\\\hline\vspace{0.1cm}
SAPCA&$\mathbf{{35.89}/{0.9439}}$ & ${\mathbf{31.35}}/{0.8671}$  & ${26.44}/{0.6587}$  & ${\mathbf{24.5}}/{0.5152}$  & $\mathbf{23.7}/{0.4465}$  & ${22.17}/{0.3273}$  & ${21.48}/{0.2783}$
\\\hline\vspace{0.1cm}
NCSR&${35.78}/{0.9375}$ & ${31.2}/{0.8551}$  & ${26.29 }/{0.6389}$  & ${24.31}/{0.4814}$  & ${23.54}/{ 0.4169}$  & ${22.02}/{0.2832}$  & ${21.36}/{ 0.2344}$
\\\hline\vspace{0.1cm}
cptTP-$\mathbb{C}$TF$_6$ &${35.56}/{0.9388}$ & ${30.94}/{0.8575}$  & ${26}/{0.6362}$  & ${24.01}/{0.4803}$  & ${23.22}/{0.4067}$  & ${21.78}/{0.2744}$  & ${21.17}/{0.2263}$ \\\hline\vspace{0.1cm}
DAS-2& ${35.44}/{{0.9373}}$  & ${30.76}/{{0.8573}}$  & ${25.91}/{0.6609}$  & ${24.03}/{0.5287}$  & ${23.22}/{0.4625}$
 & ${21.66}/{0.3316}$&${20.94}/{0.2768}$
\\\hline
\vspace{0.1cm}
BM3D&${{{35.78}}}/{0.9415}$ & ${{31.17}}/{0.8597}$  & ${{26.22}}/{0.641}$  & ${24.3}/{0.5007}$  & ${{23.58}}/{0.4286}$  & ${{22.22}}/{0.3168}$  & ${\mathbf{21.61}}/{0.2716}$ \\\hline
\vspace{0.1cm}
cbWNNM&${35.77}/{0.9412}$ & ${31.28}/{\mathbf{0.869}}$  & $\mathbf{26.45}/{{0.6838}}$  & ${{24.49}}/{0.5587}$  & ${{{23.68}}/{0.4963}}$  & ${22.05}/{0.3717}$  & ${{21.37}}/{0.3158}$ \\\hline
\vspace{0.1cm}
Hybrid&${{35.45}}/{0.9362}$ & ${31.12}/{{0.8643}}$  & ${26.23}/{\mathbf{0.684}}$  & ${24.24}/{\mathbf{0.5659}}$  & ${23.32}/{\mathbf{0.5033}}$  & ${21.57}/{\mathbf{0.3822}}$  & ${20.77}/{\mathbf{0.3292}}$ \\\hline
 \multicolumn{8}{|c|}{Man}\\
    \hline
    noised &${34.16}/{0.75}$ & ${28.14}/{0.55}$  & ${20.18}/{0.27}$  & ${16.09}/{0.16}$  & ${14.16}/{0.12}$  & ${10.07}/{0.06}$  & ${8.14}/{0.04}$ \\\hline
\vspace{0.1cm}
WNNM&${{37.99}/{ 0.8267}}$ & ${{34.19}}/{0.7234}$  & ${{29.79}}/{ 0.5403}$  & ${{27.8}}/{0.4326}$  & ${{26.95}}/{0.3842}$  & $\mathbf{25.18}/{0.2911}$  & $\mathbf{24.34}/{0.2455}$\\\hline\vspace{0.1cm}
SAPCA&$\mathbf{38.09}/{ 0.8387}$ & ${\mathbf{34.28}}/{0.7325}$  & ${\mathbf{29.84}}/{ 0.5488}$  & ${\mathbf{27.86}}/{0.445}$  & ${\mathbf{26.97}}/{0.3997}$  & ${24.92}/{0.3078}$  & ${23.9}/{0.2677}$
\\\hline\vspace{0.1cm}
NCSR&${37.88}/{0.8216}$ & ${34.08}/{0.7212}$  & ${29.62 }/{0.5289}$  & ${27.58}/{0.4107}$  & ${26.7}/{ 0.3632}$  & ${24.89}/{0.2671}$  & ${24.05}/{0.2276}$
\\\hline\vspace{0.1cm}
cptTP-$\mathbb{C}$TF$_6$ &${37.44}/{0.8302}$ & ${33.59}/{0.7235}$  & ${29.14}/{0.5341}$  & ${27.14}/{0.4241}$  & ${26.25}/{0.3725}$  & ${23.99}/{0.2713}$  & ${23.73}/{0.229}$ \\\hline\vspace{0.1cm}
DAS-2& ${37.1}/{{0.7525}}$  & ${33.16}/{{0.7184}}$  & ${28.81}/{0.5374}$  & ${26.84}/{0.4326}$  & ${25.93}/{0.3815}$
 & ${23.99}/{0.2783}$&${23.06}/{0.2336}$
\\\hline
\vspace{0.1cm}
BM3D&${{37.85}}/{{0.8306}}$ & ${{34.02}}/{0.7225}$  & ${{29.65}}/{0.5429}$  & ${27.68}/{0.442}$  & ${{26.84}}/{0.3919}$  & ${{25.09}}/{0.3009}$  & ${{22.26}}/{0.2598}$ \\\hline
\vspace{0.1cm}
cbWNNM&${37.97}/{{0.8422}}$ & ${{34.19}}/{{0.7428}}$  & ${29.81}/{0.5646}$  & ${{27.81}}/{0.4622}$  & ${{{26.82}}/{0.4173}}$  & ${{{25.04}}/{0.3216}}$  & ${{{24.11}}/{0.2806}}$ \\\hline
\vspace{0.1cm}
Hybrid&${{37.75}}/\mathbf{0.8426}$ & ${34.02}/{\mathbf{0.7452}}$  & ${29.65}/{{\mathbf{0.5674}}}$  & ${27.61}/{\mathbf{0.4664}}$  & ${26.63}/{\mathbf{0.4195}}$  & ${24.84}/{\mathbf{0.3257}}$  & ${23.88}/{\mathbf{0.285}}$ \\\hline
\end{tabular}
}
\end{table}
Table \ref{avepss}  provides the PSNR and SSIM values from  Tables \ref{tabbar} and \ref{tabman}, which are averaged over ten images participating in the experiments. Respective diagrams are drawn in Fig. \ref{diaPS}.
\begin{table}
\caption{{PSNR}/{SSIM} values averaged over 10   images. Boldface highlights  the highest values. Noise STD=$\sigma$.}\label{avepss}
\resizebox{\columnwidth}{!}{
  \centering
  \begin{tabular}{|c|c|c|c|c|c|c|c|}
    \hline
$\sigma $& 5 & 10 & 25 & 40 & 50 & 80 & 100\\\hline
 \hline
WNNM&${\mathbf{37.63}}/{0.8649}$ & ${\mathbf{33.81}}/{0.768}$  & ${\mathbf{29.46}}/{0.5949}$  & ${{27.34}}/{0.489}$  & $\mathbf{26.56}/{0.4405}$  & $\mathbf{24.8}/{0.3405}$  & $\mathbf{23.95}/{0.2916}$
\\\hline\vspace{0.1cm}
SAPCA&${{37.57}}/{0.8685}$ & ${{33.8}}/{0.7745}$  & ${{29.44}}/{0.6048}$  & ${{27.4}}/{0.5038}$  & ${26.45}/{0.4511}$  & ${24.29}/{0.3466}$  & ${23.29}/{0.3}$
\\\hline\vspace{0.1cm}
NCSR&${37.44}/{0.8586}$ & ${33.65}/{0.765}$  & ${29.2}/{0.5819}$  & ${27.13}/{0.4664}$  & ${ 26.2}/{0.4163}$  & ${24.33}/{0.3105}$  & ${23.48}/{0.2662}$
\\\hline\vspace{0.1cm}
cptTP-$\mathbb{C}$TF$_6$ &${37.03}/{0.8586}$ & ${33.2}/{0.7596}$  & ${28.72}/{0.578}$  & ${26.69}/{0.4688}$  & ${25.78}/{0.4163}$  & ${23.96}/{0.311}$  & ${23.23}/{0.2669}$
\\\hline\vspace{0.1cm}
DAS-2& ${36.98}/{{0.8565}}$  & ${33.02}/{0.7668}$  & ${28.56}/{0.597}$  & ${26.54}/{0.494}$  & ${25.61}/{0.4432}$
 & ${23.67}/{0.3376}$&${22.74}/{0.2874}$
\\\hline
\vspace{0.1cm}
BM3D&${37.4}/{0.8613}$ & ${{33.59}}/{0.7652}$  & ${{29.28}}/{0.5969}$  & ${27.15}/{0.4928}$  & ${{26.36}}/{0.4414}$  & ${{24.54}}/{0.3432}$  & ${{23.54}}/{0.2993}$ \\\hline
\vspace{0.1cm}
cbWNNM&$37.54/{\mathbf{0.8692}}$ & ${33.74}/0.7848$  & $\mathbf{29.46}/{0.6266}$ & ${\mathbf{27.43}}/{0.5317}$ & ${{{26.5}}/{0.4831}}$  & ${{24.56}}/{0.3841}$  & ${{{23.6}}/{0.3373}}$ \\\hline
\vspace{0.1cm}
Hybrid&${{37.25}}/{0.8668}$ & ${33.65}/{\mathbf{0.7854}}$  & ${29.31}/{\mathbf{0.6275}}$  & ${27.28}/{\mathbf{0.5326}}$  & ${26.37}/{\mathbf{0.4852}}$  & ${{24.46}}/{\mathbf{0.3868}}$  & ${23.53}/{\mathbf{0.3421}}$ \\\hline
\hline

\end{tabular}
}
\end{table}

\begin{figure}
\centering
\includegraphics[width=6in]{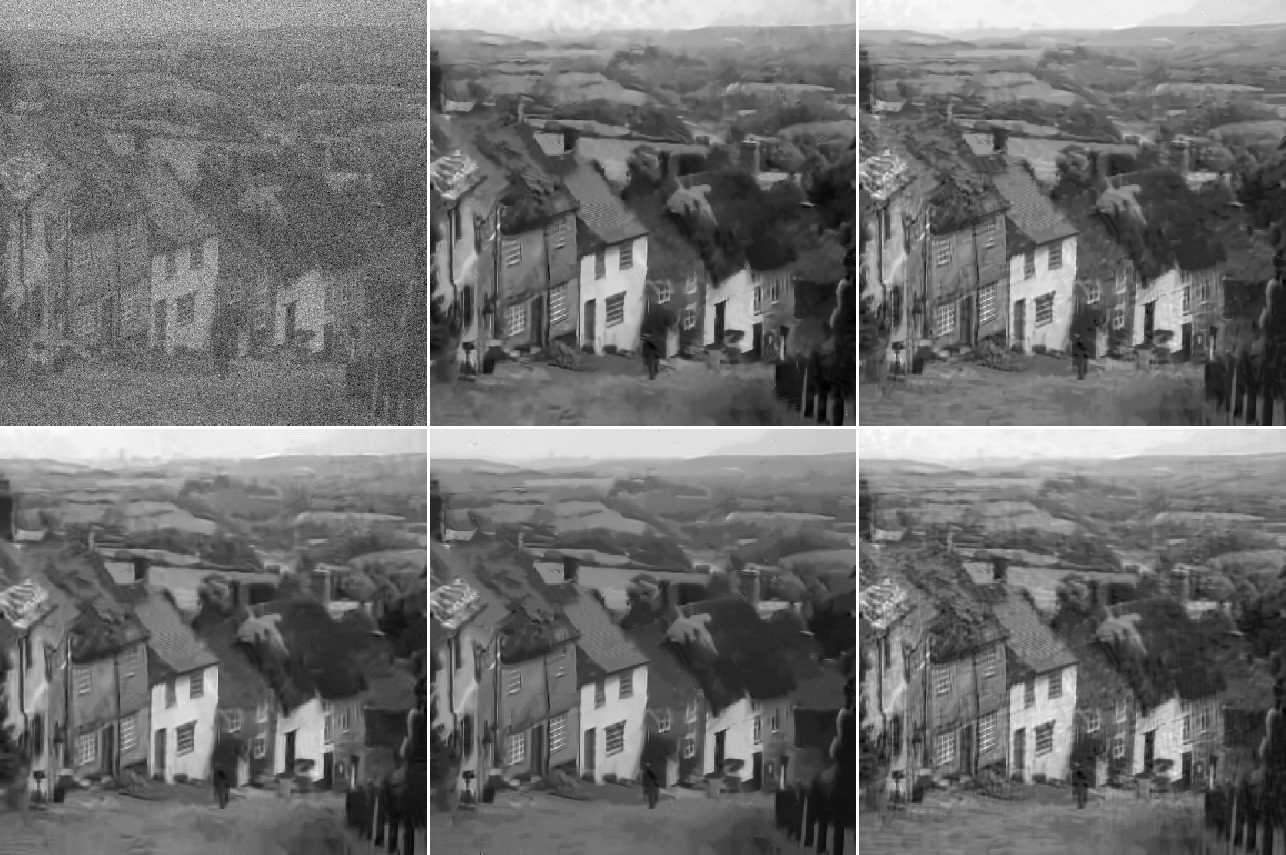}
\includegraphics[width=6in]{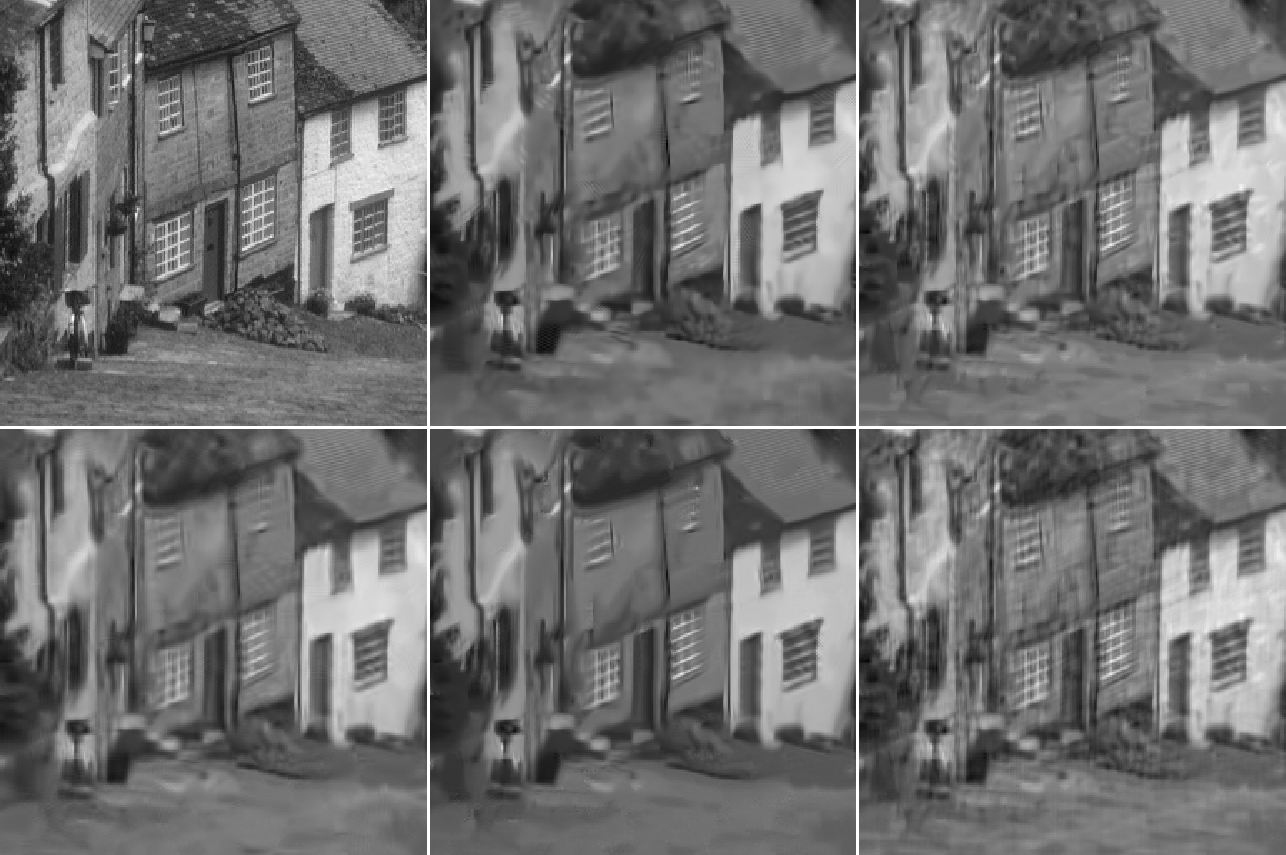}
\caption{Restoration of  the ``Hill" image corrupted by Gaussian noise with STD $\o=50$ dB. PSNR/SSIM for {BM3D}-- 27.19/0.3825; for {WNNM}-- {27.34}/0.373; for {BM3D-SAPCA}--27.2/0.3815;  for \textbf{cbWNNM}--27.4/0.4033; for  \textbf{hybrid}--27.29/{0.4118}}
\label{hill50_6}
\end{figure}

\begin{figure}
\centering
\includegraphics[width=6.0in]{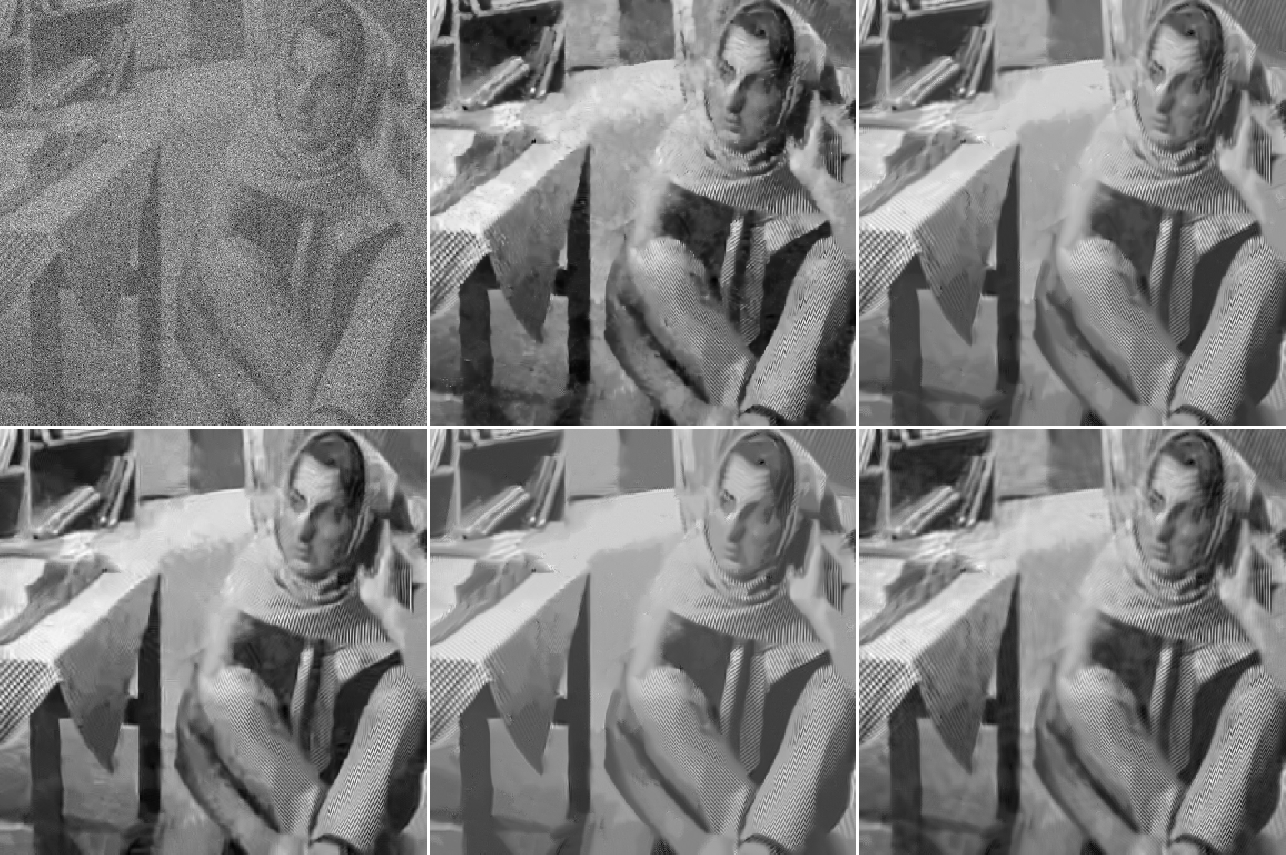}
\includegraphics[width=6.0in]{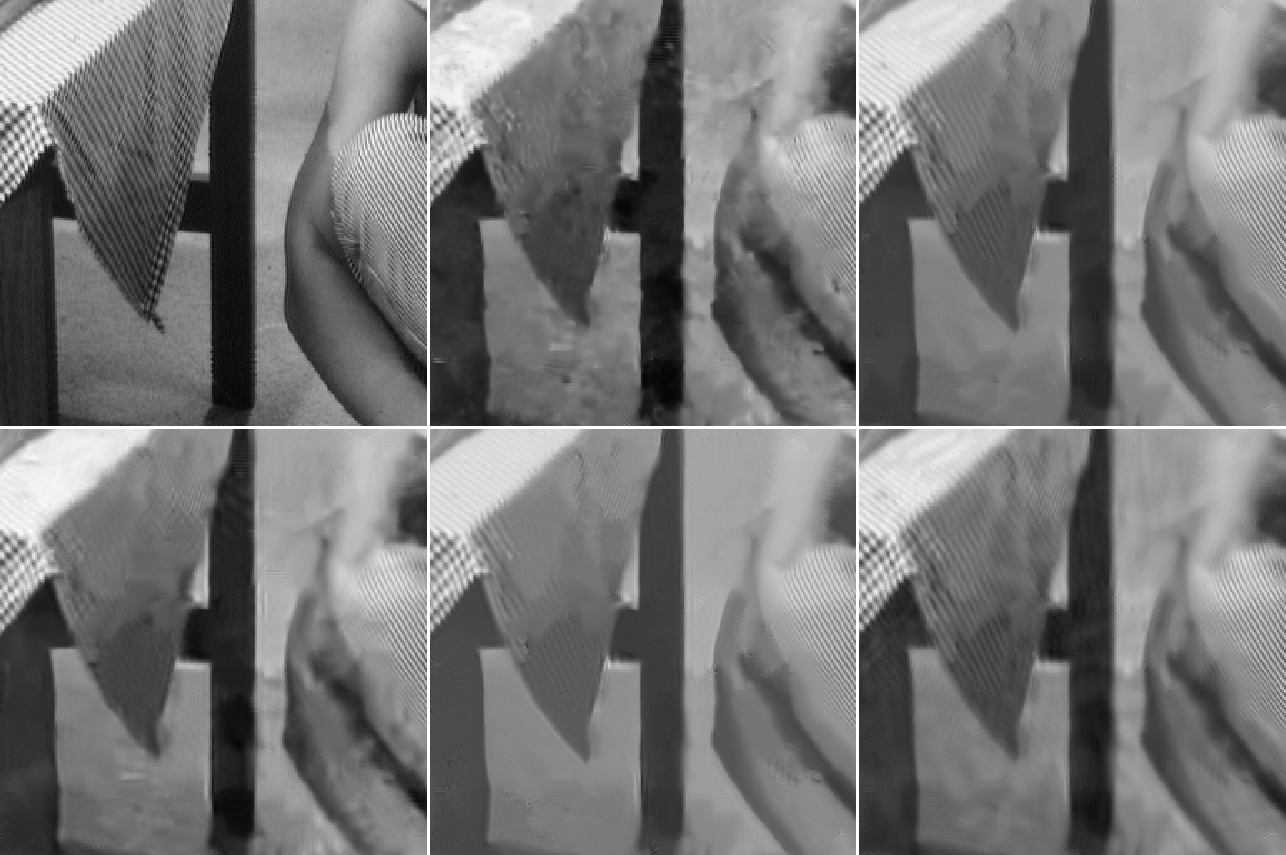}
\caption{Restoration of ``Barbara" image corrupted by Gaussian noise with STD $\o=80$ dB. PSNR/SSIM for {BM3D}-- 24.79/0.4192; for {WNNM}-- 25.48/0.4499; for {BM3D-SAPCA}--24.33/0.402;  for \textbf{cbWMMN}--25.51/{0.4681}; for  \textbf{hybrid}--{25.35}/0.4688}
\label{barb80_6}
\end{figure}

\begin{figure}
\centering
\includegraphics[width=6.0in]{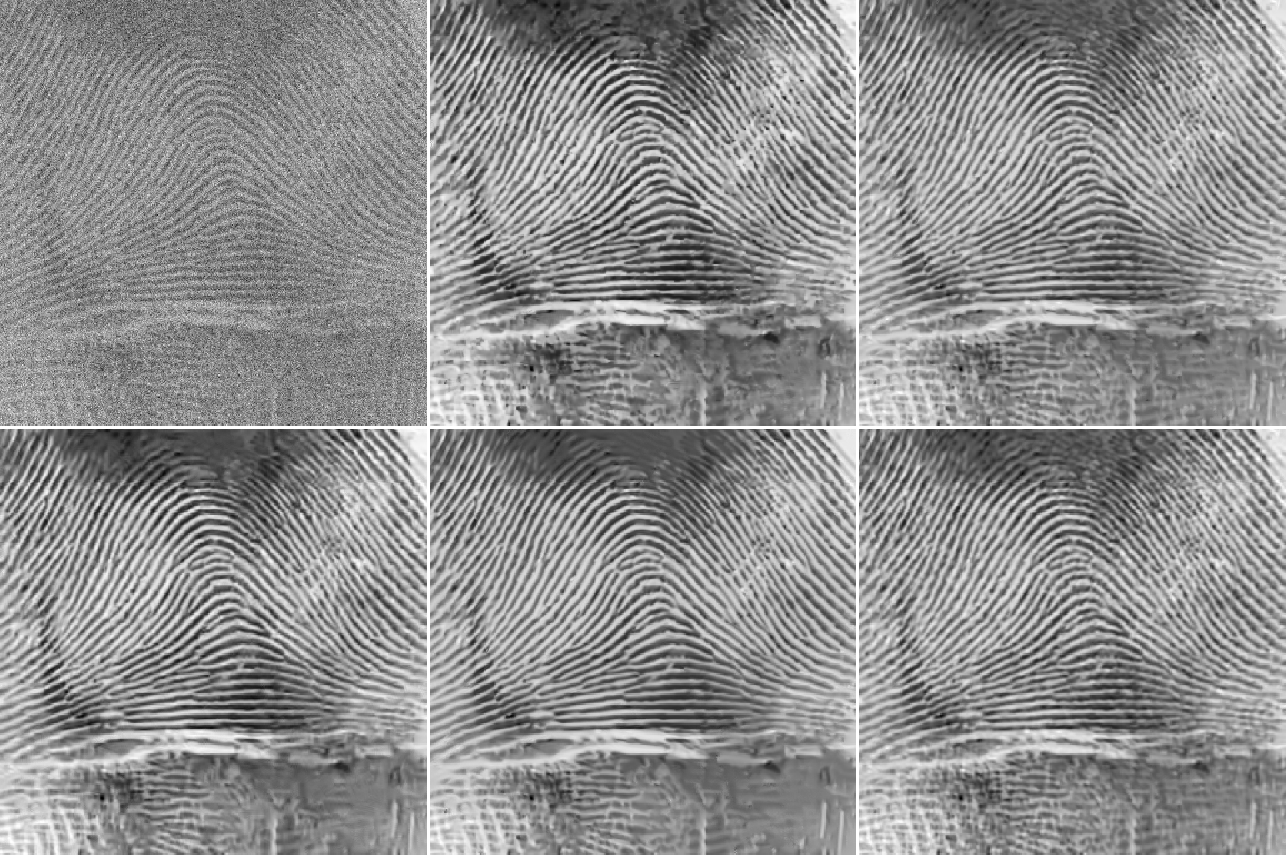}
\includegraphics[width=6.0in]{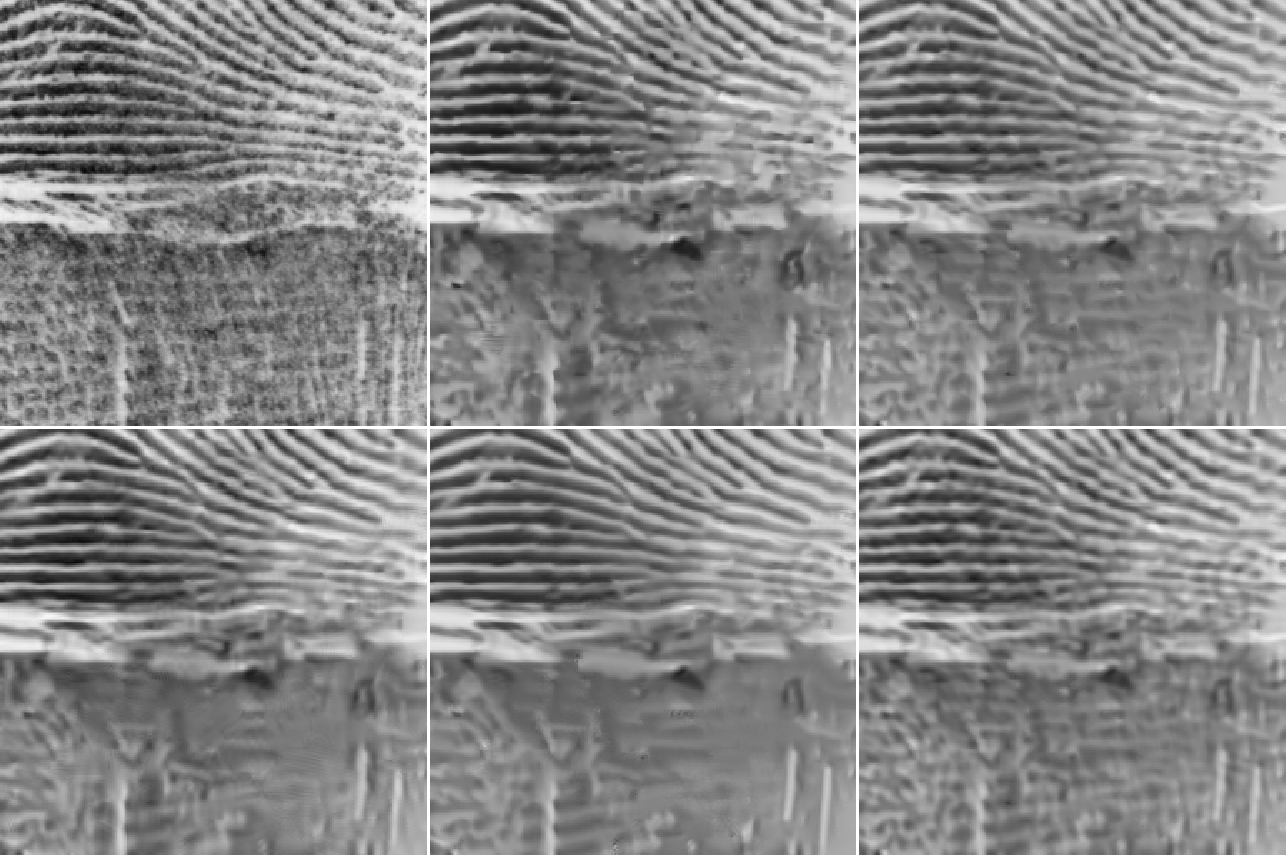}
\caption{Restoration of ``Fingerprint" image corrupted by Gaussian noise with STD $\o=80$ dB. PSNR/SSIM for {BM3D}-- 22.2/0.7165; for {WNNM}-- {22.77}/0.7289; for {BM3D-SAPCA}--22.3/0.7068;  for {upWMMN}--22.79/0.7417; for  {hybrid}--22.73/{0.7425}}
\label{fing80_6}
\end{figure}

\begin{figure}
\centering
\includegraphics[width=6in]{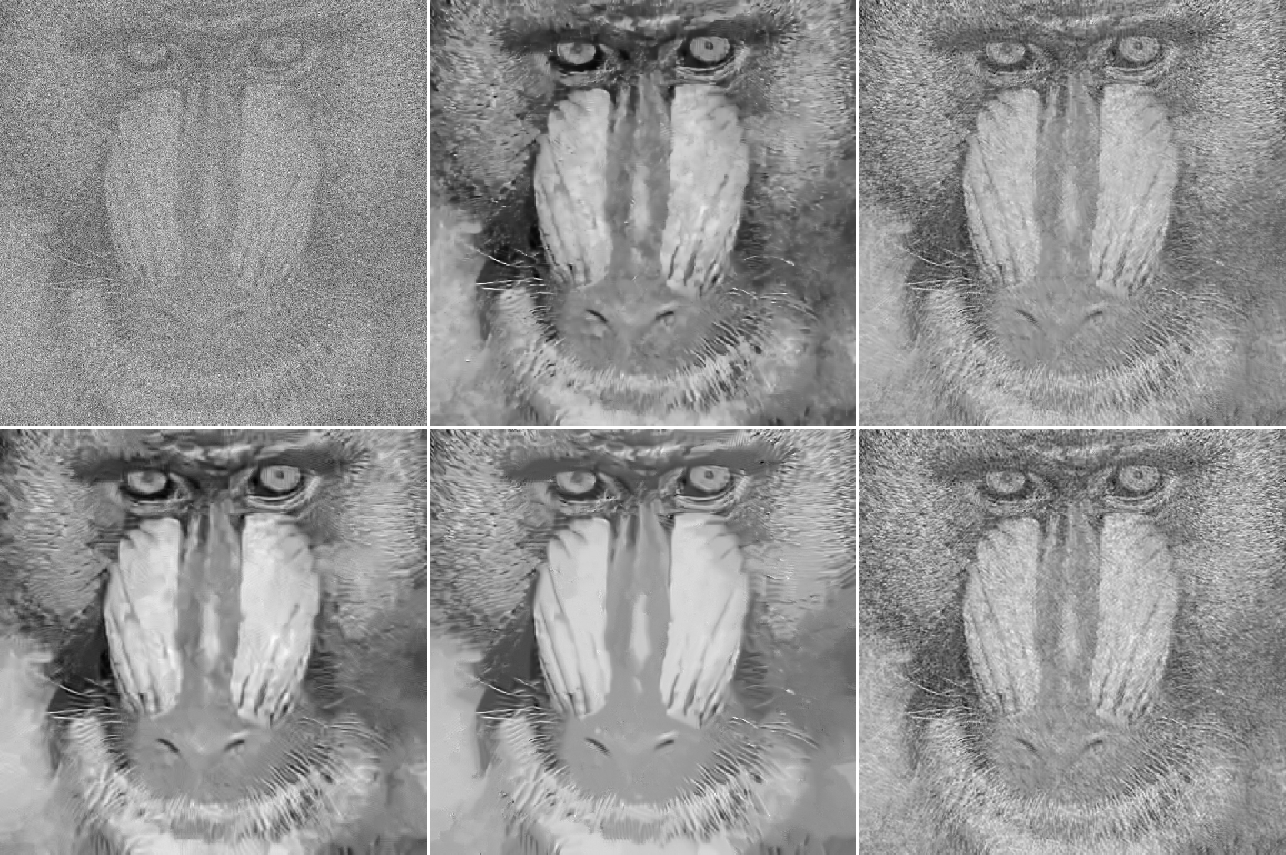}
\vspace{10pt}
\centering
\includegraphics[width=6in]{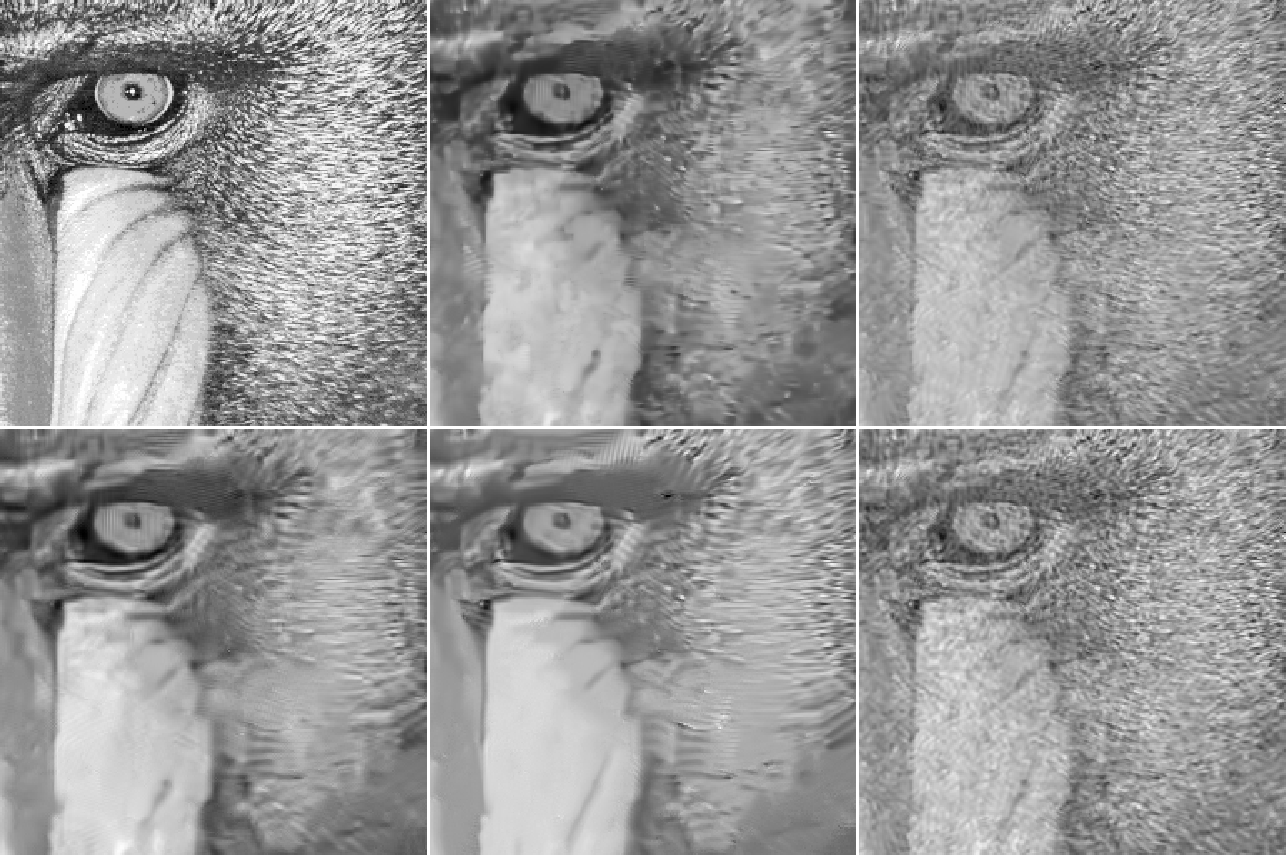}
\caption{Restoration of ``Mandrill" image corrupted by Gaussian noise with STD $\o=80$ dB. PSNR/SSIM for {BM3D}-- 20.9/0.2439; for {WNNM}-- {21.1}/0.2645; for {BM3D-SAPCA}--20.92/0.2649;  for \textbf{cbWNNM}--20.68/0.3513; for  \textbf{hybrid}--20.31/{0.3565}}
\label{mand80_6}
\end{figure}

\begin{figure}
\centering
\includegraphics[width=6.0in]{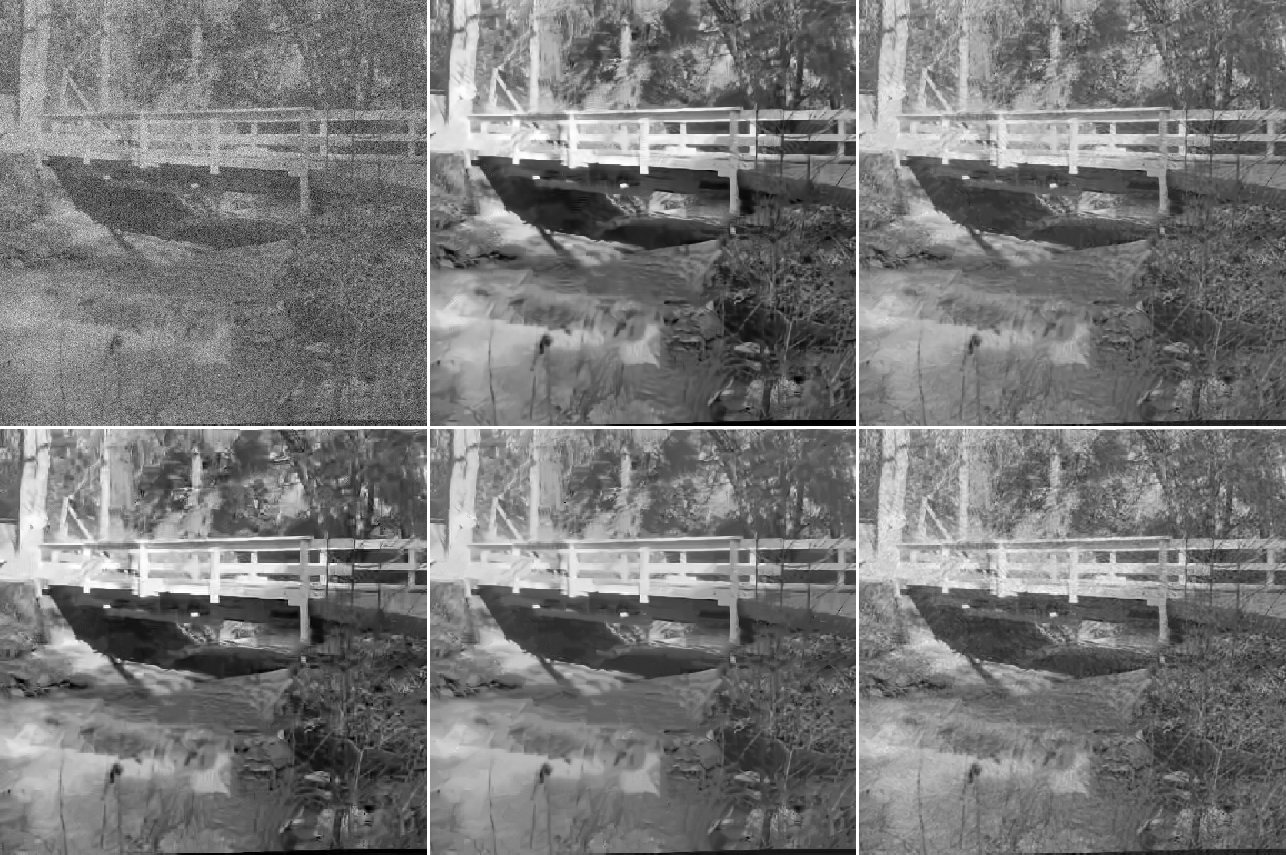}
\includegraphics[width=6.0in]{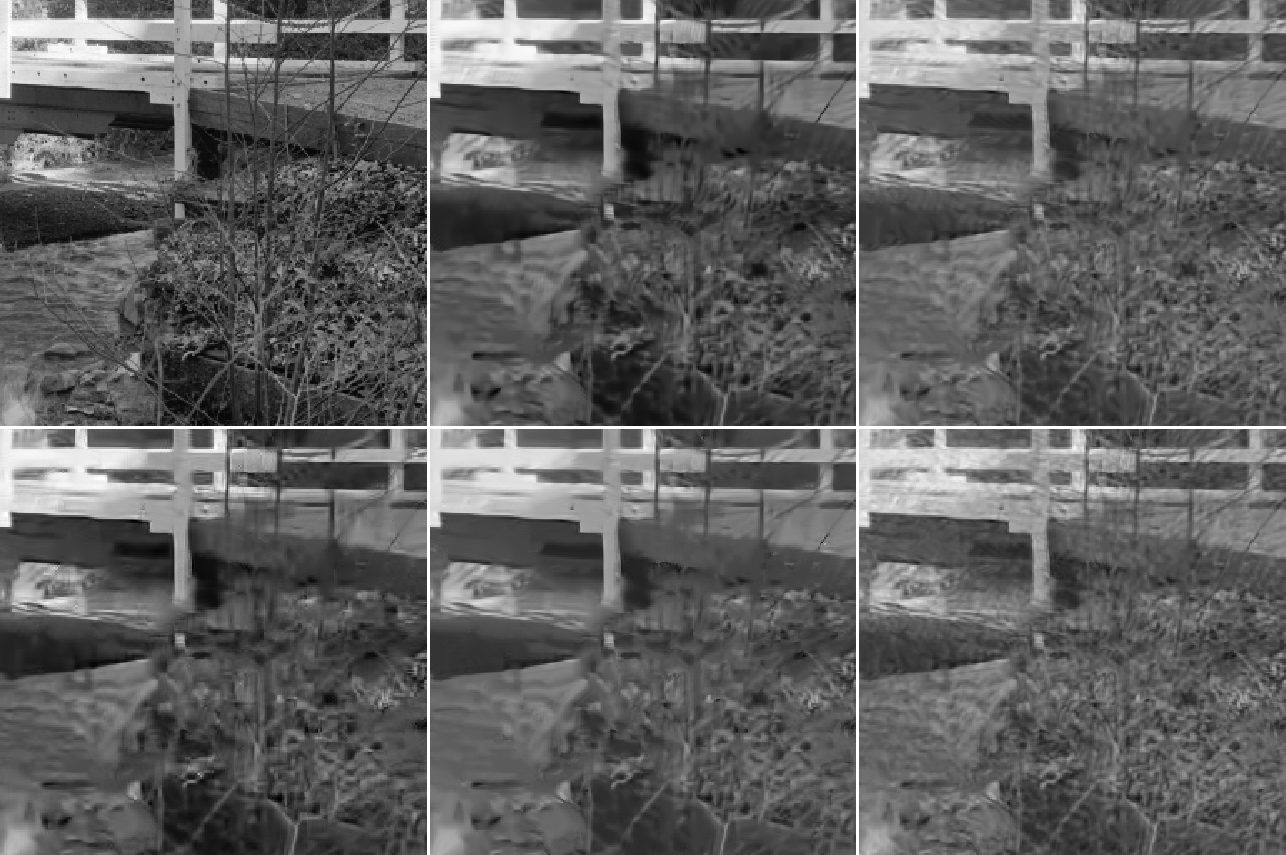}
\caption{Restoration of ``Bridge" image corrupted by Gaussian noise with STD $\o=40$ dB. PSNR/SSIM for {BM3D}-- 24.3/0.5007; for {WNNM}-- 24.45/0.4992; for {BM3D-SAPCA}--{24.5}/0.5152;  for \textbf{cbWNNM}--24.49/{0.5587}; for  \textbf{hybrid}--24.24/0.5659}
\label{bridge40_6}
\end{figure}

It is  seen from Tables \ref{tabbar}, \ref{tabman} and \ref{avepss} that our methods \textbf{cbWNNM} and {\textbf{hybrid}} produce PSNR values very close  to (sometimes higher than) those produced by the WNNM and   BM3D-SAPCA methods and higher than those from  {BM3D}, {NCSR},   cptTP-$\mathbb{C}TF_{6}$ and{ DAS-2} \dn\ \al s.  On the other hand, the SSIM values for the images restored  by   \textbf{cbWNNM} and {\textbf{hybrid}} significantly exceed  the SSIM values from all other methods. Especially it is true for the restoration of  images in presence of a strong noise. This observation reflects the fact that directional qWPs have exceptional capabilities to capture  fine structures even in  severely degraded images. This fact is illustrated  by Figures \ref{hill50_6},  \ref{barb80_6},  \ref{fing80_6},  \ref{mand80_6},  \ref{bridge40_6} and  \ref{seis25_6}, which display results of restoration of several  images corrupted by strong Gaussian noise. In all those cases the SSIM values from the \textbf{cbWNNM} and \textbf{hybrid} \al s significantly exceed values from all other \al s. Respectively, the restoration of the images' fine structure by the \textbf{cbWNNM} and \textbf{hybrid}  \al s is much better compared to the  restoration by other \al s.

Each of the mentioned figures comprises 12 frames, which are arranged in a $4\times3$ order: $\left(
                                                                                               \begin{array}{ccc}
                                                                                                 f_{11} &  f_{12} &  f_{13}  \\
                                                                                                 f_{21} &  f_{22} &  f_{23} \\
                                                                                                  f_{31} &  f_{32} &  f_{33} \\
                                                                                                 f_{41} &  f_{42} &  f_{43} \\
                                                                                               \end{array}
                                                                                             \right).$
                                                                                              Here  frame $ f_{11} $ displays noised image; frame $ f_{21} $ -- image restored by {BM3D};  $ f_{12} $ -- image restored by {BM3D-SAPCA}; $ f_{22} $ -- image restored by {WNNM};  $ f_{13} $ -- image restored by  \textbf{cbWNNM}\footnote{For the ``Seismic" image \textbf{cbqWP} instead of \textbf{cbWNNM}. }; $ f_{23} $ -- image restored by \textbf{hybrid}. Frame $ f_{31} $ displays a fragment of the original  image. The remaining frames $ \left\{ f_{32},f_{33}, f_{41} ,  f_{42} , f_{43}\right\} $ display the fragments of the restored images  shown in frames $ \left\{ f_{12},f_{13}, f_{21} ,  f_{22} , f_{23}\right\} $,  which are arranged in the same order.


\subsection{Relation of the proposed \al s to the Deep Learning methods}\label{sec:ss34}

In  recent years, the focus of the image \dn\ research shifted to the Deep Learning methods, which resulted in a huge amount of publications. We mention a few of them: \cite{tnrd,dncnn,Cruz_foi,deep_edge,Hier_res_DN,detail,complex,flash_cnn}. Much more references can be found in the reviews \cite{deep_review,dn_cnn_rev}.
One of advantages of  the Deep Learning methods is that, once the Neural Net is trained (which can involve extended datasets and take several days), its application to test images is very fast. Therefore, the experimental results in most related publications  are presented via the PSNR and, sometimes, the SSIM  values averaged over some test datasets such as, for example, Set12  introduced in \cite{dncnn}.

Set12 partially overlap with the set of 10 (Set10) images that we used in our experiments. Namely the images ``Barbara", Boat", ``Fingerprint",  ``Hill", ``Lena" and ``Man" participate in both sets. The structure of the remaining images ``Seismic", ``Fabric", ``Mandrill" and ``Bridge" from Set10 is more complicated  compared to the images
``Camera",  ``Couple",  ``House",  ``Monarch",  ``Pepper" and ``Straw" from Set12. Therefore, the averaged results comparison from these two datasets is quite justified. For even better compatibility, we compare the gains of results from \df t methods over the corresponding results from BM3D: $P_{method} \srr \frac{PSNR_{method}}{PSNR_{BM3D}}$ and $ S_{method} \srr \frac{SSIM_{method}}{SSIM_{BM3D}}$. Recall that for the calculation of SSIM we use the \fd\ \texttt{ssim} from Matlab 2020b, whereas in most publications SSIM is computed by some other schemes.

We compare results from the recent state-of-the-art \al s Cola\_Net (\cite{cola_net},  (2022)),  CDNet (\cite{complex},  (2021)), FLCNN  (\cite{flash_cnn},  (2021)), DRCNN  (\cite{detail},  (2020)), and the non-Deep Learning \al\ presented in  \cite{nlSS}   (2021), which we mark as SRENSS, averaged over Set12 with the results from WNNM \cite{wnnm} and the proposed  \textbf{hybrid} \al\ averaged over Set10\footnote{Averaged results from \textbf{upBM3D} are almost identical to those from \textbf{hybrid} \al .} The PSNR and SSIM values for all methods except for WNNM  and  \textbf{hybrid} are taken from the corresponding publications. Table \ref{aveDL} shows the results of this comparison.

\begin{table}
\caption{$P_{method}/S_{method}$, where  the PSNR and SSIM values are averaged over either Set12 or Set10.  Noise STD=$\sigma$.}\label{aveDL}
\resizebox{\columnwidth}{!}{
\begin{tabular}{|c|c|c|c|c|c|c|c|}
  \hline
  $\sigma$ & 10 & 15& 25 & 50& 70 & 80 & 100  \\
  \hline
  \hline
  Cola\_Net  & - & 1.028/1.016& 1.031/1.025 & 1.04/1.046& 1.041/1.06 & - & -  \\  \hline
  CDNet  & - & 1.019/1.016& 1.016/1.01 & 1.025/1.074 & 1.027/1.057& 1.027/1.065 & -  \\  \hline
 FLCNN  & - & 1.022/1.02 & 1.017/1.011 & 1.028/1.039 & - & - & -  \\  \hline
  DRCNN & - & 1.016/1.036& 1.02/1.033 & 1.02/1.039 & - & - &-  \\  \hline
  SRENSS& 1.044/1.011 & -& - & 1.02/1.028& 1.011/1.028 & - & 1.006/1.0038  \\  \hline\hline
  WNNM& 1.007/1.004& -  & 1.006/0.997 & 1.007/0.998& -  &1.01/0.992 & 1.018/0.974  \\  \hline
 \textbf{hybrid}& 1.002/1.026 & - &1.008/1.051  & 1.001/1.099& -  & 0.997/1.127 & 0.999/1.143 \\  \hline
  \hline
\end{tabular}
}
\end{table}
We can observe from the table that all participated up-to-date schemes, including the non-Deep Learning \al\ SRENSS, demonstrate a moderate gain over BM3D in both PSNR and SSIM  values  averaged over Set12 (far from being a breakthrough). The values from WNNM averaged over Set10 are very close to those from BM3D averaged over the same set. The same can be said  for the PSNR values from the \textbf{hybrid} \al . However, the SSIM values from the \textbf{hybrid} \al\   demonstrate a strong gain over BM3D, which is clearly seen in  Fig. \ref{diaPS} and Table \ref{avepss}. Especially it is true for a strong Gaussian noise with $\sigma=80,\,100$ dB. This fact highlights the ability of the qWP-based \al s to restore edges and fine structures even in severely damaged images.

Note that \dn\ results presented in the overwhelming majority of the Deep Learning publications  deal with the noise level not exceeding 50 dB.

\section{Discussion}\label{sec:s4}
 We presented  a \dn\ scheme that combines the  qWPdn \al\ based on the directional quasi-\az\ \wq s, which are designed in \cite{azn_pswq}, with the state-of-the-art \dn\  \al\ WNNM (\cite{wnnm})  considered to be one of the best in the field. Either of the two \al s has their strong features and shortcomings. The  qWPdn \al\ described in Section \ref{sec:ss31} demonstrates the  ability  to restore edges and texture details even from severely degraded images.  This ability stems from the fact that the designed 2D qWP \t s provide a variety of 2D testing \we s, which are close to windowed cosine waves with multiple frequencies  oriented in multiple directions. In most separate experiments, the qWPdn method demonstrated better resolution of edges and fine structures compared to the  {WNNM } \al , which were reflected by the higher SSIM values. In turn, the {WNNM } \al\  is superior for noise suppression, especially  in smooth regions in images, thus producing the highest PSNR values in almost all the experiments. However, some  over-smoothing effect on  the edges and fine texture  persisted with the WNNM \al\ when noise was strong.

 The
qWPdn and {WNNM } methods complement each other.
Therefore, the idea to combine these methods is natural. In the iterative hybrid scheme qWPdn--WNNM, which is    proposed in Section \ref{sec:ss32},  the output from one \al\ updates the input to the other. Such a hybrid  method  has some distant relation to the SOS boosting scheme presented in \cite{ela_boost}. The main distinction between  the qWPdn--WNNM  and  the SOS boosting is that each of the qWPdn and WNNM \al s is ``boosted" by the output from the other \al. Such a scheme can be regarded as a \emph{Cross-Boosting}.

The scheme  proved to be highly efficient. It is confirmed by a series of experiments on restoration of 10 multimedia images of various structure, which were degraded by Gaussian noise of various intensity. In the experiments, the performance of two combined qWPdn--WNNM \al s was compared with the performance of six advanced \dn\ \al s {BM3D}, {BM3D-SAPCA} (\cite{sapca}), {WNNM} (\cite{wnnm}),  {NCSR} (\cite{ncsr}), cptTP-$\mathbb{C}$TF$_{6}$ (\cite{Zhu_han}) and {DAS-2}  (\cite{che_zhuang}). In almost all  the experiments reported in Section \ref{sec:ss33}, the two combined \al s produce  PSNR values, which are   very close to the values produced by  { WNNM} and  {BM3D-SAPCA}. Their noise suppression efficiency is competitive with that of  { WNNM} and  {BM3D-SAPCA}.  On the other hand, their results in the resolution of edges and fine structures are much better than the  results from  all  other \al s participating in the experiments. This is seen in the images presented in  Section \ref{sec:ss33}. Consequently, the SSIM values produced by the cross-boosted \al s  qWPdn--WNNM are significantly higher than the values produced  by all other participated \al s.

Disscussion in Section \ref{sec:ss34} shows that the qWPdn--WNNM \al s can, in some aspects, compete with the Deep Learning \dn\ methods.

\pa{Acknowledgment}
This research was supported by the Israel Science Foundation (ISF,
1556/17, 1873/21), Len Blavatnik and the Blavatnik Family Foundation,
Israel Ministry of Science Technology and Space 3-16414 and  3-14481.
\bibliographystyle{plain}
   \bibliography{BookBib_TBSIA5}
\end{document}